\documentclass[11pt]{article}
\usepackage{amsfonts}
\usepackage{graphics,amssymb,amsthm,amsmath,epsf,rotating}
\usepackage{graphicx}
\usepackage{subfigure}
\usepackage{color}
\usepackage[numbers,square]{natbib}

\newtheorem{theorem}{\bf Theorem}
\newtheorem{proposition}{\bf Proposition}

\newtheorem{corollary}{\bf Corollary}
\newtheorem{definition}{\bf Definition}
\newtheorem{remark}{\bf Remark}
\newtheorem{example}{\bf Example }

\begin{document}
%%%%%%%%%%%%%%%%%%%%%%%%%%%%%%

\title{{\color{red}Measuring Inaccuracies in the Proportional Hazard Rate Model based on Extropy using a Length-Biased Weighted Residual approach } }

\author{M. Hashempour$^{1}$,
 M. R. Kazemi$^2$\thanks{kazemimr88@gmail.com} \\
%EndAName
{\small $^{1}$Department of Statistics, University of Hormozgan, Bandar Abbas, Iran}\\
{\small $^{2}$Department of Statistics, Faculty of Science, Fasa
University, Fass, Iran}
}
\date{}
\maketitle
%%%%%%%%%%%%%%%%%%%%%%%%%%%%%%%%%%%%%%%%%%%%%%
\begin{abstract}
In this paper, we consider the concept of the residual inaccuracy measure and extend it to its weighted version based on extropy. Properties of this measure are studied and the discrimination principle is applied in the class of proportional hazard rate (PHR) models. A characterization problem for the proposed weighted extropy-inaccuracy measure is studied. 
We propose some alternative expressions of weighted residual measure of inaccuracy. 
Additionally, we establish upper and lower limits and various inequalities related to the weighted residual inaccuracy measure using extropy.
Non-parametric estimators based on the kernel density estimation method and empirical distribution function for the proposed measure are obtained and the performance of the estimators are also discussed using some simulation studies. Finally, a real dataset is applied for illustrating our new proposed measure.
In general, our study highlights the potential of the weighted residual inaccuracy measure based on extropy as a powerful tool for improving the quality and reliability of data analysis and modelling across various disciplines. Researchers and practitioners can benefit from incorporating this measure into their analytical toolkit to enhance the accuracy and effectiveness of their work. 
\end{abstract}
%%%%%%%%%%%%%%%%%%%%%%%%%%%%%%%%%%%
\section{Introduction}
The concept of entropy was {\color{red}first} proposed by {\color{red}the} physicist {\color{red}Lebowitz} \cite{Lebowitz} to express the degree of chaos in the physical system.  Later, {\color{red}Shannon} \cite{13} proposed and extended entropy to the field of information as a measure of information uncertainty. Shannon entropy represents the absolute limit on the best possible lossless compression of any  communication.  {\color{red}For additional details on this concept, refer to Cover and Thomas} \cite{Cover2006}. 

  Suppose we have two non-negative continuous random variables $X$ and $Y$, which represent the time to failure of two systems. These variables have probability density functions (PDFs) $f(x)$ and $g(x)$ respectively. Additionally, let $F(x)$ and $G(x)$ be the cumulative distribution functions (CDFs) of $X$ and $Y$, and let $\bar{F}(x)$ and $\bar{G}(x)$ be the survival functions (SFs) of $X$ and $Y$ respectively. The measure of uncertainty associated with the random variable $X$, as defined by Shannon, is denoted by {\color{red}Equation \eqref{10} as} 
 \begin{eqnarray}\label{10}
 {H}(X) = -E_f[\log f(X)]=-\int ^{+\infty}_{-\infty} f(x) \log f(x) dx.
  \end{eqnarray}
Similarly, Kerridge's measure of inaccuracy, as cited in {\color{red}Kerridge} \cite{Kerridge}, is also denoted by {\color{red}Equation \eqref{20} as}
\begin{eqnarray}\label{20}
 {H}(X, Y) =-E_f[\log g(X)]=  -\int ^{+\infty}_{-\infty} f(x) \log g(x) dx,
 \end{eqnarray}
 where $``\log"$ represents the natural logarithm and following the convention that $0\log 0=0$, if we have $g(x) = f(x)$, then Equation \eqref{20} reduces to Equation \eqref{10}.

In life testing and survival analysis, the consideration of the current age of the system is important. Therefore, when calculating the uncertainty of a system or distinguishing between two systems, the measures referenced in Equations \eqref{10} and \eqref{20} may not be appropriate. Instead, when a system has survived up to time t, the corresponding dynamic measure of uncertainty of {\color{red}Ebrahimi} \cite{9} and of discrimination of {\color{red}Kayal et al. \cite{10} and Kerridge \cite{Kerridge}} are denoted by {\color{red}Equations \eqref{H(f; t)} and 
\eqref{H(f | g; t) } as}
\begin{eqnarray}\label{H(f; t)}
H(X; t)   = -\int^\infty_t   \frac{f(x)}{\bar{F}(t)} \log  \frac{f(x)}{\bar{F}(t)} dx,
\end{eqnarray}
and 
\begin{eqnarray}\label{H(f | g; t) }
    H(X| Y; t) = \int^\infty_t   \frac{f(x)}{\bar{F}(t)} \log  \frac{f(x) \bar{G}(t)}{\bar{F}(t)g(x)} dx,
\end{eqnarray}
respectively. Obviously, when $ t = 0$, then Equation \eqref{H(f; t)} reduces to Equation \eqref{10}.

According to {\color{red}Taneja et al.} \cite{225}, the dynamic measure of inaccuracy is defined as the measure associated with two residual lifetime distributions, denoted by $F$ and $G$. This measure is known as Kerridge's measure of inaccuracy and is represented by {\color{red}Equation \eqref{H(f,g ; t)}}
\begin{eqnarray}\label{H(f,g ; t)}
 H(X,Y ; t) = -\int^\infty_t   \frac{f(x)}{\bar{F}(t)} \log  \frac{g(x)}{\bar{G}(t)} dx.
\end{eqnarray}
Clearly for $t = 0$, it reduces to Equation \eqref{20}.

The relationship between information and inaccuracy can be quantified using the equation $H(X, Y) = H(X) + H(X|Y)$, where $H(X|Y)$ represents the Kullback-Leibler relative information measure of $X$ about $Y$ of {\color{red}Kullback} \cite{Kullback}, defined {\color{red}in Equation \eqref{3000}} as
\begin{eqnarray}\label{3000}
 {H}(X|Y) = E_f \left[\log \frac{f(X)}{g(X)}\right]= \int ^{\infty}_0 f(x) \log \frac{f(x)}{g(x)}dx.
\end{eqnarray}
It is clear that $H(X| Y; 0)=H(X| Y)$.

The information measures mentioned in the {\color{red}Di Crescenzo and Longobardi} 
\cite{2000} do not consider the value of the random variable itself, but only its probability density function (PDF). {\color{red}They} proposed a ``length-biased'' shift dependent information measure that is related to the differential entropy. This measure assigns higher weight to larger values of the observed random variables.

The concept of weighted distribution, as introduced by {\color{red}Rao} \cite{11}, is widely utilized in statistics and various other applications. Weighted distributions come into play when observations generated from a stochastic process are recorded with a certain weight function. In this context, let $X$ represent a non-negative continuous random variable with a PDF $f(x)$. Additionally, let $X_w$ be a weighted random variable associated with $X$, where the weight function $w(x)$ is positive for all values of $x\geq 0$. The corresponding PDF $f_w(x)$ of $X_w$ can be determined {\color{red}in Equation \eqref{w1}} as 
\begin{eqnarray}\label{w1}
f^w(x)=\dfrac{w(x) f(x)}{E\left(w\left( X \right) \right)}, \quad x \geq 0.
\end{eqnarray}

When the function $w(x) = x$ is used, the random variable $X_w$ is referred to as a length biased or size biased random variable. In this case, the PDF becomes
\begin{eqnarray}\label{w2}
f_*(x)=\dfrac{x f(x)}{E\left(X \right)}, \quad x \geq 0.
\end{eqnarray}

{\color{red}For additional details on this subject, refer to} {\color{red}Patil and Ord} \cite{Patil} and {\color{red}Furman and Zitikis} \cite{Furman}. If $X$ is a random variable with a finite mean $E[X]$, the length biased CDF and SF are defined {\color{red}in Equations \eqref{w3} and \eqref{w4}} as
\begin{eqnarray}\label{w3}
F_*(t)= \int_0^t \dfrac{x f(x)}{E\left(X \right)}dx, 
\end{eqnarray}
and 
\begin{eqnarray}\label{w4}
\bar{F}_*(t)= \int_t^\infty \dfrac{x f(x)}{E\left(X \right)}dx, 
\end{eqnarray}
respectively. 
These functions describe weighted distributions that occur in sampling procedures where the probabilities of sampling are proportional to the values of the samples. 
Therefore, the measure of residual entropy in Equation  \eqref{H(f; t)} has been expanded to include the length biased weighted residual entropy, denoted by {\color{red}Equation \eqref{eq-shi-ref-1} as} 
\begin{equation}\label{eq-shi-ref-1}
H_*(X, t)=-\int_t^{\infty} x \frac{f(x)}{\bar{F}(t)} \log \frac{f(x)}{\bar{F}(t)} d x .
\end{equation}
The presence of the factor $x$ in the integral on the right-hand side introduces a length biased shift-dependent information measure that assigns higher significance to larger values of the random variable $X$.
{\color{red}Di Crescenzo and Longobardi} \cite{Di 2006} discussed weighted versions of the residual and past entropies. The weighted residual entropy is defined {\color{red}in Equation \eqref{eq-shi-ref-2}} as
\begin{equation}\label{eq-shi-ref-2}
H^w\left(X_t\right)=-\int_t^{+\infty} x \frac{f(x)}{\bar{F}(t)} \log \frac{f(x)}{\bar{F}(t)} \mathrm{d} x,
\end{equation}
while the weighted past entropy is defined {\color{red}in Equation \eqref{eq-shi-ref-3}} as
\begin{equation}\label{eq-shi-ref-3}
H^w\left({ }_t X\right)=-\int_0^t x \frac{f(x)}{F(t)} \log \frac{f(x)}{F(t)} \mathrm{d} x .
\end{equation}
% % % % % % % % % % % % %
\section{Extropy}
\noindent  Recently, {\color{red}Lad et al.} \cite{1000}  proposed an alternative measure of uncertainty of a random variable called extropy. 
The extropy is a measure of information introduced as dual to entropy or  as an antonym to entropy. The extropy of the random variable $X$ is defined {\color{red}in Equation \eqref{eq1.1ex}} as 
\begin{eqnarray}\label{eq1.1ex}
 J(X) =-\frac{1}{2}\int ^{\infty}_{0} f^2(x) dx  
 =  -\frac{1}{2}\int ^{+\infty}_{0} f(x) dF(x)
= -\frac{1}{2}\int^1_0 {f \big(F ^{-1}( u)\big)}du.
 \end{eqnarray}
For more details and applications of extropy, {\color{red}refer to Lad et al. \cite{1000}}.
 Extropy may also be used to compare the uncertainties of two random variables. For two random variables $Y$ and $Z$, $J(Y) \leq J(Z)$ implies that $Y$ has more uncertainty than $Z$.   
  
{\color{red}Qiu and Jia} \cite{1004} considered a random variable $ X_t = [X -t|X > t],~ t\geq 0$ and defined uncertainty  of such a system based on extropy, given {\color{red}in Equation \eqref{H(f; t)J0}} as
\begin{eqnarray}\label{H(f; t)J0}
 J(X; t)=-  \frac{1}{2}  \int^\infty_t  \left[ \frac{f(x)}{\bar{F}(t)} \right]^2 dx.
\end{eqnarray}
    
Analogous to the weighted entropy, {\color{red}Balakrishnan et al.} \cite{bala-et-al-22} introduced the concept of weighted extropy {\color{red}defined in Equation \eqref{eq-bala-et-al-22}} as
\begin{eqnarray} \label{eq-bala-et-al-22}
J^w(X)=-\frac{1}{2} E[X f(X)]=-\frac{1}{2} \int_0^{+\infty} x f^2(x) \mathrm{d} x,
\end{eqnarray}
which can also be rewritten {\color{red}in Equation \eqref{eq-shi-ref-5}} as
\begin{eqnarray}\label{eq-shi-ref-5}
J^w(X)=-\frac{1}{2} \int_0^{\infty} f^2(x) \int_0^x \mathrm{~d} y \mathrm{~d} x=-\frac{1}{2} \int_0^{\infty} \mathrm{d} y \int_y^{\infty} f^2(x) \mathrm{d} x .
\end{eqnarray}
Also, they introduced the weighted residual extropy {\color{red}by Equation \eqref{eq-shi-ref-6}} as
 \begin{eqnarray}\label{eq-shi-ref-6}
J^w\left(X_t\right)=-\frac{1}{2 \bar{F}^2(t)} \int_t^{\infty} x f^2(x) \mathrm{d} x.
\end{eqnarray}
{\color{red}Jahanshahi et al.} \cite{{1010}} introduced an alternative measure of uncertainty of non-negative continuous random variable $X$
which they called it cumulative residual extropy (CRJ) {\color{red}by Equation \eqref{eq1.1exx}} as
\begin{eqnarray}\label{eq1.1exx}
 \xi J(X) =-\frac{1}{2}\int ^{\infty}_{0} \bar{F}^2(x) dx.
 \end{eqnarray}
They studied some properties of the aforementioned information measure. 
The measure defined in Equation \eqref{eq1.1exx} is not applicable to a system which has survived for some unit of time. Hence, {\color{red}Sathar and Nair} \cite{sath} proposed a dynamic version of CRJ (called dynamic survival extropy) to measure residual uncertainty of lifetime random variable $X$ as follows
\begin{equation}\label{exp000} 
\xi J(X; t) =-\frac{1}{2\bar{F}^2(t)}\int_{t}^{\infty}
\bar{F}^2 (x)dx, \ \ t\geq 0.
\end{equation}
 It is clear that $\xi J(X;0) = \xi J(X)$.
Recently, {\color{red}Hashempour et al.} \cite{Hashempour} introduced a weighted cumulative residual extropy as an extended version of {\color{red}Equation} \eqref{eq1.1exx} as follows
 \begin{eqnarray}\label{H}
 \xi J^w(X) =-\frac{1}{2}\int ^{\infty}_{0} x \bar{F}^2(x) dx.
 \end{eqnarray}
 They studied the characterization problem, estimation and testing for this measure. 
 Also, {\color{red}Mohammadi and Hashempour} \cite{Mohammadi-Hashempour-2} proposed a modified interval weighted cumulative residual and past extropies, respectively in {\color{red}Equations} \eqref{IWCRJ} and \eqref{IWCPJ} as
\begin{equation}\label{IWCRJ}
WCRJ(X; t_1, t_2) =-\frac{1}{2}\int^{t_2}_{t_1} \varphi(x) \left( \frac{\bar{F} (x)}{\bar{F} (t_1)-\bar{F} (t_2)} \right)^2
 dx,
\end{equation}
and
\begin{equation}\label{IWCPJ}
WCPJ(X; t_1, t_2) =-\frac{1}{2}\int^{t_2}_{t_1} \varphi(x) \left( \frac{F (x)}{F (t_2)-F (t_1)} \right)^2  dx,
\end{equation}
where $\varphi$ is a weight function.
%Measures in equations \eqref{IWCRJ} and \eqref{IWCPJ} are  generalized measures of equations \eqref{exp000} and \eqref{DCPJ1}.
They provided non-parametric estimators for these measures based on the kernel method.
Also, For more details, concepts, generalizations, applications and estimations in the field of extropy, one can {\color{red}refer to} the following references.
{\color{red}Qiu} \cite{2017} used the extropy for record values and ordered statistics.
{\color{red}Qiu and Jia} \cite{2018a} considered the estimations of extropy measure.
The extropy of a mixed system's lifetime was considered by {\color{red}Qiu and Jia} \cite{2019}. 
Recently, {\color{red}Hashempour and Mohammadi} \cite{hash-moh-ext-rec-2024} consider the extropy measure of inaccuracy for record statistics. 
Also,  readers can refer to {\color{red}Kazemi et al.} \cite{kazemi}, {\color{red}Hashempour et al.} \cite{Hashempour}, {\color{red}Hashempour and Mohammadi} \cite{Hashempour-Mohammadi-1},    
{\color{red}Pakdaman and Hashempour} \cite{P1,P2}, and references therein.
In this article, an attempt has been made to present new criteria that have a more general form compared to the criteria introduced in other articles. One of these criteria is the presentation of residual weight criteria based on extropy. In this regard, data available often do not have equal importance and value. In such cases, the presented criteria should be considered in a weighted form. Additionally, in many cases, there is a need to obtain information about future events, where we use the SF instead of the PDF. Considering the aforementioned points, the extropy-based residual weight criteria can meet the needs of researchers.
In this paper we extend the concept of residual inaccuracy to length biased weighted residual inaccuracy. In the rest of this paper, In Section $3$, we define and study the weighted measure of inaccuracy  and weighted discrimination information based on extropy. In Section $4$, we study that when $F$ and $G$ follow the PHR model then dynamic residual measure of inaccuracy uniquely determines the survival function $\bar{F}$ and propose some alternative expressions of weighted measure of inaccuracy and some properties of this measure are studied. In Section $5$, we obtain some bounds and inequalities for our proposed measure. In Section $6$, Two non-parametric estimators for the proposed measure as well as a simulation study are also obtained. Finally, Section $7$ investigates the behavior of the proposed estimators for a real dataset.
We conclude the paper in Section \ref{sec-conclusion}.
\section{Weighted  residual inaccuracy measure } \label{se1.3}
In this section,  for two non-negative continuous random variables with the same support,  we introduce some  measures of uncertainty based on extropy and some properties are studied.

\definition
\rm Let $X$ and $Y$ be two non-negative continuous random variables with PDFs  $f$ and $g$, respectively. The 
weighted measure of discrimination of $X$ about $Y$ based on extropy is defined {\color{red}in Equation \eqref{kullbackJ}} as
\begin{equation}\label{kullbackJ}
J^w(X | Y)=\frac{1}{2}  \int^\infty_0 x f(x) \left[ f(x)-g(x) \right]dx.
\end{equation}
%By adding equations \eqref{shanonJ} and \eqref{kullbackJ}, 
The weighted measure of inaccuracy between $X$ and $Y$ based on extropy denoted by $ J^w(X,Y) $ is defined {\color{red}in Definition \eqref{def-shi-ref-1}} as follows.
%we define a new measure of inaccuracy as denoted by $ J(X, Y) $ is defined as follows.

\definition \label{def-shi-ref-1}
\rm Let $X$ and $Y$ be non-negative continuous random variables with PDFs $f(x)$ and $g(x)$ and CDFs $F(x)$ and $G(x)$, respectively. Then the weighted measure of inaccuracy between the distributions $X$ and $Y$ is defined {\color{red}in Equation \eqref{sum1}} as
[Weighted  Extropy-Inaccuracy (WJI)]
\begin{eqnarray}\label{sum1}
 J^w(X,Y) =-\frac{1}{2}\int ^{\infty}_{0} x f(x) g(x) dx.
\end{eqnarray}
\rm On WJI measure in Equation \eqref{sum1}, $\bar {F}(.)$ is the actual SF corresponding to the observations and $\bar {G}(.)$ is the SF assigned by the experimenter. If $f(x)=g(x)$ then WJI in Equation \eqref{sum1} reduces to Equation \eqref{eq-bala-et-al-22} introduced by {\color{red}Balakrishnan et al.} \cite{bala-et-al-22}. WJI measures the value of mis-specifying the correct model in which $f(x)$ is the actual PDF of observations and $g(x)$ is the PDF assigned by experimenter such that the value of each observation is taken into account in its formula. In other words, by WJI, the measure of discrimination between $f$ and $g$ can be affected by the strength of each observation.  

In the provided example, we demonstrate the application of Equation \eqref{sum1} for the purpose of comparing statistical models.
\example \label{ex1}
\rm The statistical model for the random variable $X$ is represented by the SF $\bar{Q}(x)=1-x$, where $x$ is within the range $(0,1)$. Additionally, two SFs, $\bar{F}(x)=1-x^2$  and $\bar{S}(x)=1-x^3$, $	 x \in (0, 1)$, have been determined through non-parametric statistical tests to approximate the random variable $X$. Using Equation \eqref{sum1}, we can calculate the WJI  values. Specifically, we have $J^w(X,X)=J^w(X)=-0.25$, $J^w(X,Y)=-0.33$ and $J^w(X,Z)=-0.5$. Based on these WJI values, we can conclude that the WJI between X and Y, which follows the SF $\bar{F}(x)$, is closer to the WJI of $X$ itself compared to the WJI between $X$ and $Z$, which follows the SF $\bar{S}(x)$. Therefore, $Y$ provides a better approximation to X than $Z$. In other words, the statistical model represented by the survival function $\bar{F}(x)$ is the closest approximation to the statistical model represented by $\bar{Q}(x)$ that generated the data.  \hfill{$\Box$}\\

 %Qiu and Jia \cite{1004} 
Balakrishnan et al. \cite{bala-et-al-22} examined a random variable denoted by $ X_t = [X -t|X > t]$, where $t\geq 0$. They introduced the concept of uncertainty in this system using extropy {\color{red}in Equation \eqref{H(f; t)J}}, denoted by
\begin{eqnarray}\label{H(f; t)J}
 J^w(X; t)=-  \frac{1}{2}  \int^\infty_t x \left[ \frac{f(x)}{\bar{F}(t)} \right]^2 dx.
\end{eqnarray}
  The weighted residual extropy $J^w(X;t)$ is a suitable metric for quantifying information in situations where uncertainty is linked to future events.
  
In the subsequent discussion, we introduce a weighted measure of inaccuracy that pertains to two residual lifetime distributions, denoted by $G(.)$ and $F(.)$, which are associated with the measure of inaccuracy.
\definition
\rm  Consider two non-negative continuous random variables $X$ and $Y$ with probability density functions $f$ and $g$ respectively. The weighted residual inaccuracy (WRJI) measure between $X$ and $Y$, utilizing extropy, can be defined {\color{red}in Equation \eqref{H(f,g ; t)J}} as follows
   \begin{eqnarray}\label{H(f,g ; t)J}
 J^w(X,Y;t) =-\frac{1}{2} \int^\infty_t  x  \frac{f(x)}{\bar{F}(t)}\frac{g(x)}{\bar{G}(t)} dx,
 \end{eqnarray}
 meanwhile, $\bar{F}(t)$ and $\bar{G}(t)$ can not be zero in \eqref{H(f,g ; t)J}.
 
\begin{remark}\label{R1}
\rm From equation\eqref{H(f,g ; t)J}, it is observed that $J^w(X,Y;t) = J^w(Y,X;t)$ and $J^w(X,Y;t) \leq 0$. Additionally, by taking the limit as $t \rightarrow 0$ in Equation \eqref{H(f,g ; t)J}, the WRJI transforms into the measure of inaccuracy in Equation \eqref{sum1}. Furthermore, when two random variables $X$ and $Y$ have the same SFs, the WRJI simplifies to the WRJ as given in Equation \eqref{H(f; t)J}.
% Indeed, closer the value of WRJI is to the weighted dynamic residual extropy, the better $Y$ is an approximation of $X$.
%The proof is easy and hence omitted for brevity.
\end{remark}

\begin{remark}\label{Rem-weight}
\rm The weight functions used in the weighted residual measure of inaccuracy play a crucial role in determining the impact of different data points on the overall measure of model fit. It is important to explain why these particular weight functions were chosen over others and how they align with the objectives of the study. Discussing the rationale behind the choice of weight functions and their potential effects on the results will enhance the understanding of the methodology used in the study. Additionally, sensitivity analysis on different weight functions could be beneficial to assess the robustness of the results.
\end{remark}

In addition, we introduce the concept of uncertainty weighted discrimination information of variable $X$ regarding variable $Y$, utilizing extropy.

\definition
\rm Assume that $\bar{F}(x)$ and $\bar{G}(x)$ denote the SFs of non-negative continuous random variables $X$ and $Y$, respectively. The weighted residual discrimination information (WRDJ) between $X$ and $Y$ can be defined {\color{red}in Equation \eqref{H(f | g; t)J }} as a quantity denoted by 
\begin{eqnarray}\label{H(f | g; t)J }
J(X| Y; t) =  \frac{1}{2} \int^\infty_t x  \frac{f(x)}{\bar{F}(t)}\left[ \frac{f(x)}{\bar{F}(t)}-\frac{g(x)}{\bar{G}(t)}\right] dx.
\end{eqnarray}
Clearly when $t = 0$, then Equations \eqref{H(f; t)J}-\eqref{H(f | g; t)J } reduce to  Equations \eqref{eq1.1ex}, \eqref{sum1} and \eqref{kullbackJ}, respectively. By adding Equations \eqref{H(f; t)J}  and \eqref{H(f | g; t)J }, we obtain  Equation \eqref{H(f,g ; t)J}, i.e.  
$J^w(X,Y;t)=J^w(X;t)+J^w(X|Y;t) $.\\
The WRJI measure based on extropy offers a valuable approach for evaluating predictive models in a dynamic data environment. By incorporating weights and extropy, this measure provides a more comprehensive assessment of model performance, considering the importance of different time periods and the unpredictability of the data. The practical value of WRJI lies in its ability to capture the nuances of data and provide a more accurate evaluation of predictive models. This measure can be applied in various fields where accurate predictions are crucial, such as finance, health care, weather forecasting, and supply chain management. By considering the weighted residuals and extropy, decision-makers can gain insights into the model's performance over time and make informed decisions based on the most relevant and reliable information. Furthermore, the potential impact of WRJI extends beyond model evaluation. It can aid in model selection, parameter tuning, and identifying areas for improvement in predictive models. By understanding the strengths and weaknesses of different models in dynamic scenarios, organizations can enhance their decision-making processes and optimize their operations. Altogether, the WRJI measure based on extropy offers a robust and practical solution for evaluating predictive models in dynamic environments. Its potential impact in diverse fields is significant, enabling more accurate predictions, informed decision-making, and improved performance of predictive models.
 
\example
\rm Let $X$ be a non-negative random variable with SF of $\bar{F}(x)= 1- x^2$ for values of $x$ between 0 and 1. Furthermore, $Y$ is a random variable with a uniform distribution between 0 and 1, and its SF is denoted by  $  \bar{Q}(x)=1-x$, $x \in (0, 1).$
we obtain 
$$ J^w(X, Y; t)=\dfrac{t^2+t+1}{3t^2-3}, ~~~  t \neq 1,$$
$$ J^w(X, t)= \dfrac{t^2+1}{2t^2-2} ,$$
and
$$ J^w(X|Y; t)=-\dfrac{t-1}{6t+6}, ~~~  t \neq 1.$$

%Figure $\ref{}$ provides the graphs of $ J^w(X, Y; t)$ and $ J^w(X|Y; t) $ for various values of $t$ in the case where $X$ and $Y$ are random variables with the given SFs.

%I checked so far  @@@@@@@      %%%%%%%%%%%%%%%
\example \label{_exp_dist}
\rm Suppose that $X$ and $Y$ have exponential distributions with SFs as follows
\begin{align*}
&\bar{F}(t)=e^{-\theta t}; \qquad  t\geq 0, \ \theta>0, \\
&\bar{G}(t)=e^{-\lambda t}; \qquad  t\geq 0, \ \lambda>0.
\end{align*}
From Equation \eqref{sum1}, we given
\begin{align*}
  J^w(X, Y)= -\dfrac{{\theta}{\lambda}{e^{t(\theta+\lambda)}}}{2\left({\lambda}+{\theta}\right)^2},
\end{align*}
\begin{align*}
  J^w(X, Y;t)=   -\dfrac{{\theta}{\lambda}\cdot\left(t{\lambda}+t{\theta}+1\right)}{2\left({\lambda}+{\theta}\right)^2{e^{-t(\theta+\lambda)}}}.
\end{align*}

%In this case,  $J(X, Y)=J(X, Y; t)$, because the exponential distribution has a memoryless feature.
%Functions $J^w(X, Y)$ and $ J^w(X, Y; t)$ are shown in Figure \ref{} for some selected values of $\theta$ and $\lambda$.
%It is obvious that $J^w(X, Y)$ is a \textcolor{blue}{ non decreasing} function in terms of $\theta$ and $\lambda$.\hfill{$\Box$}
%Also, it can be seen that $J^w(X, Y; t)$ is a constant function with respect to $ t $.\hfill{$\Box$}

\example \label{weib_dist}
\rm Let $X$ and $Y$ have Weibull distributions with same shape parameter 2 and SFs as
\begin{align*}
&\bar{F}(t)=e^{-\theta t^2}; \qquad  t\geq 0, \ \theta>0, \\
&\bar{G}(t)=e^{-\lambda t^2}; \qquad  t\geq 0, \ \lambda>0.
\end{align*}
From Equation \eqref{sum1}, we obtain
\begin{align*}
  J^w(X,Y)= -\dfrac{{\theta}{\lambda}{e^{t^{2}(\theta+\lambda)}}}{\left({\lambda}+{\theta}\right)^2},
\end{align*}
also, from Equation \eqref{H(f,g ; t)J}, we have
% \textcolor{blue}{
\begin{align*}
   J^w(X,Y;t)=-\dfrac{{\theta}{\lambda}\left(t^{2}({\lambda}+{\theta})+1\right)}{2\left({\lambda}+{\theta}\right)^2{e^{-t^{2}(\theta+\lambda)}}}.
 %  -\frac{\theta \lambda}{\theta + \lambda}\left( t+\frac{1}{t} \right).
\end{align*}
%}
%The behaviour of functions $J^w(X, Y)$ and $J^w(X, Y; t) $ are shown in Figure \ref{}.
%It can be seen that $J^w(X, Y; t)$ is a \textcolor{blue}{non decreasing} function in terms of $\theta$, $\lambda$, and $ t $.\hfill{$\Box$}

%$-\dfrac{\operatorname{\Gamma}\left(\frac{3}{2},\left(b+a\right)t^2\right)\,ab\mathrm{e}^{bt^2+at^2}}{\left(b+a\right)^\frac{3}{2}}
%$

In what follows, we prove another result to show the effect of monotone transformations on WJI defined in Equation \eqref{sum1}. In this context, we prove the following theorem.
\begin{theorem}
\rm Let $X$ be a non-negative absolutely continuous random variable with PDF $f(x)$ and CDF $F(x)$. Assume $Y=\phi(X)$, where $\phi$ is a strictly monotonically increasing and differentiable function.  Let $G(y)$ and $g(y)$ denote the distribution and density functions of $Y$, respectively. Then, 
%, with derivative $\xi^{'}(x)$
\begin{equation}
J^w(Y)=J(X, \frac{\phi(X)}{\phi^{'}(X)}X ). 
\end{equation}
\end{theorem}
\begin{proof}
The PDF of $Y=\phi(X)$ is $g_Y(y)=  | \frac{1}{\phi^{'}(\phi^{-1}(y))}| f_X(\phi^{-1}(y)).$ Therefore,
\begin{eqnarray*}
J^w(Y)=-\frac{1}{2}\int ^{\infty}_{0} y g^2_Y(y) dy.
%&=&   -\frac{1}{2}\int ^{\infty}_{0}       \left[ \frac{1}{\eta^{'}(\eta^{-1}(y))}\right]^2 f^2_X(\eta^{-1}(y))dy
\end{eqnarray*}
This gives
\begin{eqnarray*}
J^w(Y)= -\frac{1}{2}\int ^{\infty}_{0} y \left[ \frac{1}{\phi^{'}(\phi^{-1}(y))}\right]^2 f^2_X(\phi^{-1}(y))dy
\end{eqnarray*}
Substituting $x=\phi^{-1}(y)$, we get
\begin{eqnarray*}
J^w(Y)&=&-\frac{1}{2}\int ^{\infty}_{0}\phi(x) \left( \frac{1}{\phi^{'}(x)}\right)^2 f^2_X(x) \phi^{'}(x)  dx\\
&=& -\frac{1}{2}\int ^{\infty}_{0}   f^2_X(x) \frac{\phi(x)}{\phi^{'}(x)}    dx\\
&=& -\frac{1}{2}\int ^{\infty}_{0}f_X(x)  \frac{\phi(x)}{\phi^{'}(x)}   f_X(x) dx  \\
&=& J(X, \frac{\phi(X)}{\phi^{'}(X)}X),  
\end{eqnarray*}
 the proof is completed.        \hfill{$\Box$}
\end{proof}

\example \label{ex}
\rm Suppose $\bar{F}(x)=\exp\{- \theta x\}$, $	x \in (0, \infty)$, is the true statistical model for random variable $X$ that generated some data. Also, suppose $\bar{G}(x)=\exp\{-2 \theta x\}$ and $\bar{S}(x)=\exp\{-5 \theta x\}$, $x \in (0, \infty)$, be two SFs determined through non-parametric statistical tests to approximate $X$. From Equation \eqref{sum1}, we obtain $J^w(X,X)=J^w(X)= -1/8$, $J^w(X,Y)=-1/9$ and $J^w(X,Z)=-5/72$. Thus, WJI between $X$ and random variable $Y$ which follows the survival function $\bar{G}(x)$ is closer than that of between $X$ and random variable $Z$ which follows the survival function $\bar{S}(x)$. Therefore, $Y$ provides a better approximation to $X$ than $Z$ i.e. the statistical model $\bar{F}(x)$ is the closest to the statistical model $\bar{G}(x)$ that generated data.  
\hfill{$\Box$}\\

 \example
\rm According to Example \ref{ex} and  Equation \eqref{H(f,g ; t)J}, we have
 
 $$ J^w(X,X; t)= J(X; t)=-\dfrac{2t{\theta}+1}{8},$$
 
 $$ J^w(Y, Y; t)= J(Y; t)=-\dfrac{4t{\theta}+1}{8},$$ 
 
$$ J^w(Z, Z; t)= J( Z; t)=-\dfrac{10t{\theta}+1}{8}.$$ 
 
  Also,   from Equation \eqref{H(f,g ; t)J},  we obtain 
  
  $$ J^w(Y, Z; t)=-\dfrac{35t{\theta}+5}{49},$$
 
  $$ J^w(X, Z; t)=-\dfrac{30t{\theta}+5}{72}.$$
It is seen that $J^w(Y, Z; t)$ is greater than $ J^w(X, Z; t)$ for all $t,\theta>0$. \\
%Figure \ref{} provides the graphs  for various values of $t$ in the case where $X$, $Y$ and $Z$ are random variables with survival functions $\bar{F}(x)$,  $\bar{G}(x)$ and $\bar{S}(x)$, respectively.\hfill{$\Box$}

\example
\rm Let $X$ and $Y$ be two non-negative random variables with survival functions $\bar{F}(x) =(x+1)e^{-x}$ and $\bar{G}(x) = e^{-2x}, ~ x > 0$ respectively. We obtain
$$ J^w(X, Y; t)=-\dfrac{9t^2+6t+2}{27\left(t+1\right)},$$
and
$$ J^w(X|Y; t)=-\dfrac{36t^3+78t^2-34t-49}{432t^2+864t+432}.$$

\begin{figure}[ht]
\centering
\includegraphics[scale=0.5]{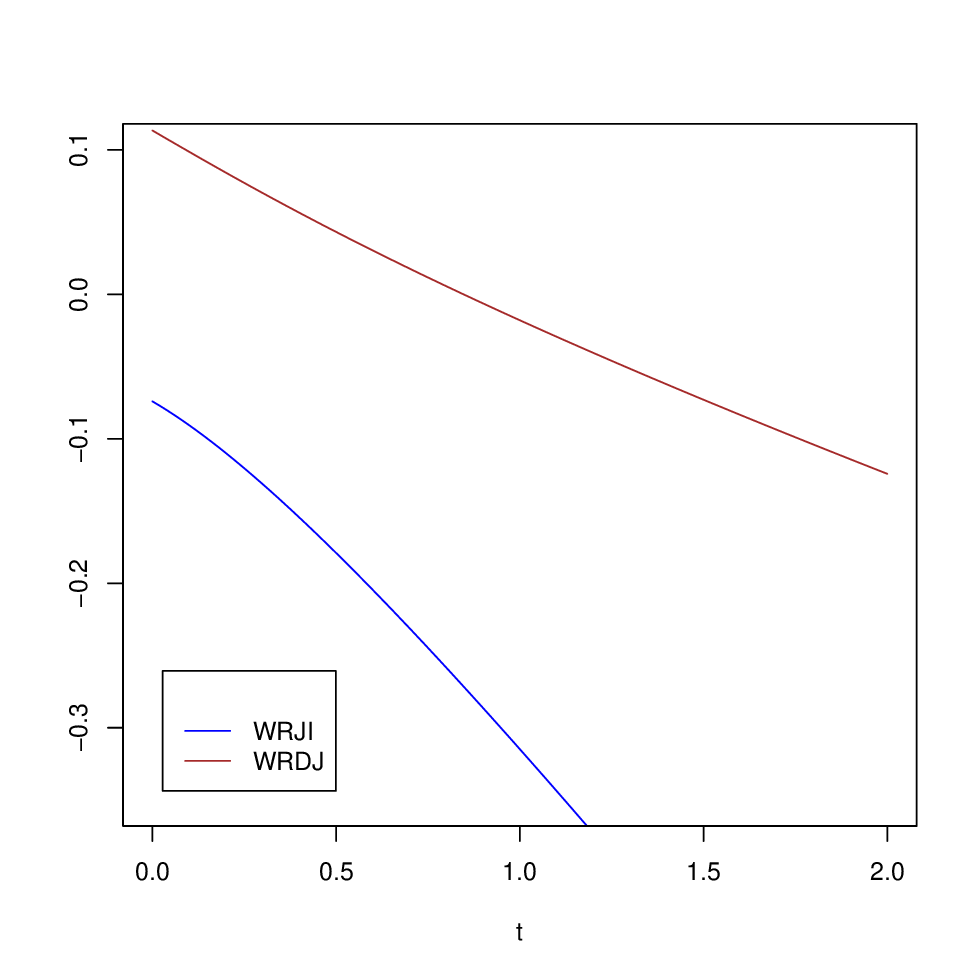}
\vspace{-.5cm}
\caption{{\color{red}Graph of $ J^w(X, Y; t)$ and $ J^w(X|Y; t)$ as a function of $t$.}}\label{fig-ex-7}
\end{figure}

Figure \ref{fig-ex-7} provides the graphs of $ J^w(X, Y; t)$ and $ J^w(X|Y; t)$ for various values of $t$ in the case where $X$ and $Y$ are random variables with SFs. From Figure \ref{fig-ex-7}, we can see that both of $ J^w(X, Y; t)$ and $ J^w(X|Y; t)$ are decreasing function of $t$. %\hfill{$\Box$}

In the following, we will investigate the relationship between WRIJ and WJI.
\begin{corollary}
\rm Suppose that  $X$ and $Y$ are non-negative continuous random variable with SFs $\bar{F}(x)$  and  $\bar{G}(x)$, respectively. Then, we given
\begin{eqnarray}
J^w(X,Y;t)=a(t) \left[ J^w(X,Y)+c(t)\right],
\end{eqnarray}
where $a(t)=[ \bar{F}(t) \bar{G}(t)]^{-1}$  and  $c(t)= \frac{1}{2} \int^t_0 x f(x)g(x) dx$.
\end{corollary}
\begin{proof}
\begin{eqnarray*}
J^w(X,Y; t)  
%&=& - \frac{1}{2} \int^\infty_t  x \frac{f(x)}{\bar{F}(t)}  \frac{g(x)}{\bar{G}(t)} dx \\
&=&  - \frac{1}{2} \left[ \int^\infty_0  x \frac{f(x)}{\bar{F}(t)}  \frac{g(x)}{\bar{G}(t)} dx  -\int^t_0 x \frac{f(x)}{\bar{F}(t)}  \frac{g(x)}{\bar{G}(t)} dx \right] \\
&=&\frac{1}{\bar{F}(t) \bar{G}(t)} \left[ J^w(X,Y) + \frac{1}{2}
\int^t_0  x f(x)g(x)  dx  \right].  
\end{eqnarray*}
This completes the proof. 
\end{proof}

%\example
%Assume that $X$ and $Y$ have Pareto distribution with SFs given by
%\begin{align*}
%&{F}(x)=1- \left(\frac{2}{x}\right)^c; \ f(x)=\frac{c2^c}{x^{c+1}}, \ \ c>1, \ x\geq 0, \\
%&{G}(t)=1- \left(\frac{2}{x}\right)^b; \ g(t)=\frac{b2^b}{x^{b+1}}, \ \  b>1, \ x\geq 0.
%\end{align*}
% After some algebraic calculations, we given
%\[ J^w(X, Y)=-\dfrac{bc}{2\left(c+b\right)},\]
%\[ J^w(X, Y; t)= -\dfrac{bc}{2\left(c+b\right)}, \]
%\[ J^w(X, t)=-\dfrac{c}{4},\]
%and
%\[ J^w(X|Y; t)=\dfrac{c\left(c-b\right)}{4\left(c+b\right)}.\]
%The functions $J^w(X, Y; t)$,  $ J^w(X, Y)$ and $J^w(X|Y; t)$  are shown in Figures  \ref{} for some selected values of $c$ and $k$.\hfill{$\Box$}
\begin{corollary}\label{28}
\rm Let  $X$ and $Y$ be two non-negative continuous random variables with PDFs  $f$ and $g$, respectively. Then, we given
\begin{eqnarray}
J^w(X,Y;t) = k_1 J^w(X,Y)-k_2 \bar{J}^w(X,Y;t),
\end{eqnarray}
where $k_1=[\bar{F}(x)\bar{G}(x)]^{-1}$, $k_2=\frac{F(x)G(x)}{\bar{F}(x)\bar{G}(x)}$ and $\bar{J}^w(X,Y;t)$ is the weighted past inaccuracy measure.
\end{corollary}

%Weighted dynamic residual inaccuracy measure
\section{Some properties of  WRJI measure}
\rm Recently, several authors studied the subject  of characterizing underlying distribution of a sample based on the extropy. 
The general characterization problem is to determine when the residual
measure characterizes the CDF uniquely. In this section, we study characterization problem for the proposed weighted residual inaccuracy measure in Equation \eqref{H(f,g ; t)J}  under the PHR model.
We study characterization problem for the WRJI under the assumption that the distribution functions of $X$ and $Y$ satisfy the PHR model. Under this model, the SFs of two random lifetime variables are related by
 \begin{eqnarray}\label{PRHRM}
\bar{G}(x) = \left[   \bar{F}(x)   \right]^\gamma, \quad \gamma>0. 
\end{eqnarray}
%  Characterization property on the measure of weighted measure of inaccuracy between   two residual life length distributions $F(.)$ and $G(.)$ corresponding to the measure of inaccuracy  is studied by using the sufficient condition for the uniqueness of the solution of initial value problem given by ........... \\
%  The  characterization problem is to determine when the dynamic measure determines the distribution function uniquely. 

Notice that, based on the PHR model in Equation \eqref{PRHRM}, the hazard rate functions (HRFs) $\mu_F (x)$ and $\mu_G(x)$ satisfy the relation $\mu_G(x) = \gamma \mu_F (x)$. 
The PHR model assumes that the hazard rate for an individual at any given time is a constant multiple of the hazard rate for a reference individual. This assumption implies that the hazard ratios between different groups remain constant over time. It is important to discuss why this assumption is appropriate for the specific context of the study and how it aligns with the data being analyzed. Justifying the use of the PHR model will add credibility to the results obtained using this methodology.
Since  $X$ and $Y$ satisfying the PHR model, the  $J^w(X,Y)$, $J^w(X,Y;t)$ and $J^w(X|Y;t)$  can rewrite  as follows.
%\begin{corollary}\label{1}

%\definition \label{1}
Let $X$ and $Y$ be two non-negative continuous random variables satisfying the PHR model. Then, we have 
%\begin{description}
%  \item[(i)]
  \begin{eqnarray}\label{J(f, g)phr}
 (I) \quad J^w(X, Y)=- \frac{\gamma}{2}\int_0^\infty x\mu^2_F (x) \bar{F}^{\gamma+1}(x)dx,
\end{eqnarray}
% \item[(ii)]
 \begin{eqnarray}\label{J(f, g; t)phr}
(II) \quad  J^w(X, Y; t) =- \frac{\gamma}{2}\int_t^\infty x\mu^2_F (x) \left[ \frac{\bar{F}(x)}{\bar{F}(t)} \right]^{\gamma+1}dx, 
\end{eqnarray}
\begin{eqnarray}\label{3phrm}
(III) \quad  J^w(X|Y;t)=\frac{1}{2}\int_0^\infty x\left( \frac{{f}(x)}{\bar{F}(t)} \right)^2
 \left[1-\gamma \left( \frac{\bar{F}(x)}{\bar{F}(t)} \right)^{\gamma+1} \right]dx.
\end{eqnarray}
 %&=&- \frac{\gamma}{2}\int_t^\infty \frac{f^2(x)\bar{F}^{\gamma-1}(x)}{\bar{F}^{\gamma+1}(t)}dx\\
%\\
%&=& -\frac{\gamma}{2}\int_t^\infty \left( \frac{\bar{F}(x)}{\bar{F}(t)} \right)^{\gamma-1}  \left( \frac{{f}(x)}{\bar{F}(t)} \right)^2  dx 
%\end{description}
 %\end{corollary}
 When $\gamma=1$, that is, $\bar{F}(x)=\bar{G}(x) $, then Equation \eqref{J(f, g; t)phr} becomes Equation \eqref{H(f; t)J}.
% \begin{proof}
% We consider
%\begin{eqnarray}\label{J(f, g; t)}
%J^w(X, Y; t)=-\frac{1}{2}\int_t^\infty \frac{f(x)g(x)}{\bar{F}(x)\bar{G}(x)}dx
%\end{eqnarray}
%Using the relation  $\mu_G(x) = \gamma \mu_F (x)$,  \eqref{J(f, g; t)}  becomes
%\begin{eqnarray}\label{J(f, g; t)phr}
%J^w(X, Y; t) &=&- \frac{\gamma}{2}\int_t^\infty x \frac{f^2(x)\bar{F}^{\gamma-1}(x)}{\bar{F}^{\gamma+1}(t)}dx\\
%&=&- \frac{\gamma}{2}\int_t^\infty x \mu^2_F (x) \left( \frac{\bar{F}(x)}{\bar{F}(t)} \right)^{\gamma+1}dx 
%\\
%&=& -\frac{\gamma}{2}\int_t^\infty x\left( \frac{\bar{F}(x)}{\bar{F}(t)} \right)^{\gamma-1}  \left( \frac{{f}(x)}{\bar{F}(t)} \right)^2  dx 
%\end{eqnarray}
%\end{proof}

Some alternative expressions to Equations \eqref{J(f, g)phr} and \eqref{J(f, g; t)phr}  of WJI and WRJI of a non-negative random variable $X$ are provided hereafter.
\begin{corollary}\label{1}
Let  $X$ and $Y$ be two non-negative continuous random variables satisfying the PHR model. Then, we have 
\begin{eqnarray}\label{J(f, g)phr2}
(I) \quad J^w(X, Y)=- \frac{\gamma}{2}\int_0^\infty xf^2(x)\bar{F}^{\gamma-1}(x) dx,
\end{eqnarray}
\begin{eqnarray}\label{J(f, g; t)phr2}
(II) \quad J^w(X, Y; t)=-\frac{\gamma}{2}\int_t^\infty x\left( \frac{\bar{F}(x)}{\bar{F}(t)} \right)^{\gamma-1}  \left( \frac{{f}(x)}{\bar{F}(t)} \right)^2  dx,
\end{eqnarray}
 \begin{eqnarray}\label{J(f, g; t)phr3}
(III) \quad J^w(X, Y; t)=- \frac{\gamma}{2}\int_t^\infty xf^2(x) \left( \frac{\bar{F}^{\gamma-1}(x)} 
{\bar{F}^{\gamma+1}(t)} \right) dx.
\end{eqnarray}
\end{corollary}
%In the following, we prove the uniqueness result.
%In this following, we show that the measure of inaccuracy WDRJI can  determine  the underlying distribution uniquely. Also, we study the following some properties of the  WDRJI.
\rm In this following, we show that the measure of inaccuracy WRJI can  determine the underlying distribution uniquely. Also, we study the following properties of the  WRJI.
\begin{theorem}\label{thm1}
\rm Let $X$ and $Y$ be two non-negative continuous random variables satisfying the PHR model, then $J^w(X,Y; t)$ uniquely determines the survival function $\bar{F}(x)$ of random variable $X$.
\end{theorem}
\begin{proof}
Suppose  $X_1$, $Y_1$ and $X_2$, $Y_2$ are two sets of random variables satisfying PHR model, that is, $\lambda_{G_1}(x) = \gamma \lambda_{F_1} (x)$, $\lambda_{G_2}(x) = \gamma \lambda_{F_2} (x)$ and
\begin{equation}\label{1=2}
J^w(X_1, Y_1; t)=J^w(X_2, Y_2; t),  \quad t \geq 0.
\end{equation}
By differentiating from both side of Equation \eqref{1=2}  with respect to $t$ and using $\lambda_G(x) = \gamma \lambda_F (x)$, we have
\begin{eqnarray}
\frac{d}{dt}J^w(X, Y; t) &=&-\frac{\gamma}{2}\left[ -t\lambda^2_F(t) +(\gamma +1)\frac{f(t)\int_t^\infty x \lambda^2_F(x)\bar{F}^{\gamma +1}(x)   dx}{\bar{F}^{\gamma +2}(t)}\right] \\ \nonumber
%&-& \frac{1}{2} \frac{g(t)\bar{F}(t)\int_t^\infty f(x)g(x)dx}{\bar{F}^2(t)\bar{G}^2(t)}\\
&=&-\frac{\gamma}{2}\left[  -t \lambda^2_F(t)+(\gamma +1  ) \lambda_F(t) \int_t^\infty x\lambda^2_F(x)\left( \frac{\bar{F}(x)}{\bar{F}(t)} \right)^{\gamma +1}dx
\right] \\ \nonumber
&=&
\frac{ \gamma}{2}t\lambda^2_F(t)+\lambda_F(t) J^w(X, Y; t) +\gamma \lambda_F(t) J^w(X, Y; t)\\ \nonumber
&=&\lambda_F(t) \left[ \frac{t\gamma \lambda_F(t)}{2} + (\gamma+1) J^w(X, Y; t)\right].
\end{eqnarray}
From Equation \eqref{1=2} we given
\begin{eqnarray}\label{11=22}
\lambda_{F_1}(t) \left( \frac{\gamma t \lambda_{F_1}(t)}{2} + (\gamma+1) J^w(X_1, Y_1; t)\right)=
\lambda_{F_2}(t) \left( \frac{\gamma t \lambda_{F_2}(t)}{2} + (\gamma+1)J^w(X_2, Y_2; t)\right).
\end{eqnarray}
Now to prove that Equation \eqref{1=2}, under the assumption of PHR model in Equation \eqref{PRHRM}, implies  $\bar{F}_1(t)=\bar{F}_2(t)$, it is sufficient to prove that
\begin{eqnarray}
\lambda_{F_1}(t)=\lambda_{F_2}(t), \quad \forall t>0.
\end{eqnarray}
In the sequel, define a set  $\Omega =\{ t: t \geq 0, and \quad \lambda_{F_1}(t) \neq  \lambda_{F_2}(t)   \}$
%\begin{equation}\label{A}
%A=\{ t: t \geq 0, and \quad \mu_{F_1}(t) \neq  \mu_{F_2}(t)   \}
%\end{equation}
and suppose the set $\Omega $ is not empty. Thus for some $t_0 \in \Omega,~  \lambda_{F_1}(t_0) \neq  \lambda_{F_2}(t_0) $. Without loss of generality assume that $\lambda_{F_1}(t_0) > \lambda_{F_2}(t_0)$ and hence Equation \eqref{11=22} holds, when either
\begin{eqnarray}\label{<}
\frac{\gamma t \lambda_{F_1}(t)}{2} + (\gamma+1) J^w(X_1, Y_1; t_0) <\frac{\gamma t \lambda_{F_2}(t)}{2} + (\gamma+1) J^w(X_2, Y_2; t_0)
\end{eqnarray}
or
\begin{eqnarray}\label{=}
\frac{\gamma t \lambda_{F_1}(t)}{2} + (\gamma+1) J^w(X_1, Y_1; t_0) =\frac{\gamma t \lambda_{F_2}(t)}{2} + (\gamma+1) J^w(X_2,Y_2; t_0)=0.
\end{eqnarray}
Let inequality \eqref{<} holds, then using Equation \eqref{1=2}, inequality  \eqref{<} reduces to
$\lambda_{F_1}(t_0) < \lambda_{F_2}(t_0) $. If equality \eqref{=} holds, then using Equation \eqref{1=2}, it reduces to $\lambda_{F_1}(t_0) = \lambda_{F_2}(t_0) $. Combining these two results, we get $\lambda_{F_1}(t_0) \leq\lambda_{F_1}(t_0) $. This contradicts our assumption and therefore  set $\Omega$ is empty and this concludes the proof. 
\end{proof}

According to Equation \eqref{H(f,g ; t)J}, the measure of inaccuracy  WRJI  has ordinary upper bound $0$. We will establish a lower bound.
\begin{remark}
\rm Let  $X$ and $Y$ be two non-negative random variables satisfying the PHR model. If  $M=f (m) \leq \infty$, where $m = \sup\{x: f (x) \leq M\}$ is the mode of $X$, then  
\begin{eqnarray}
 a_1 M^2  \leq J^w(X,Y;t) \leq 0,
\end{eqnarray}
where $a_1=-\frac{\gamma}{2\bar{F}^{\gamma+1}(t)}\int_t^\infty x \bar{F}^{\gamma-1}(x) dx.$
\end{remark}
Similarly,  according to Equation \eqref{sum1}, we have
\begin{eqnarray}
 a_2   M^2 \leq   J^w(X,Y) \leq 0,
\end{eqnarray}
where $a_2=-\frac{\gamma}{2}\int_0^\infty x \bar{F}^{\gamma-1}(x) dx.$

\begin{proposition}\label{p3}
\rm Let $X$ and $Y$ be two non-negative continuous random variables satisfying the PHR model. The maxima of weighted dynamic residual inaccuracy  WDRJI measurer  exist when $F$ is exponential.
\end{proposition}
\begin{proof}
\begin{eqnarray}\label{max}
J^w(X, Y; t) &=&- \frac{\gamma}{2\bar{F}^{\gamma+1}(t)}\int_t^\infty x f^2(x)\bar{F}^{\gamma-1}(x)dx  \nonumber \\
&=& \frac{\gamma}{2\bar{F}^{\gamma+1}(t)}\int_t^\infty x f^2(x) \left[1-1+\bar{F}^{\gamma-1}(x)\right]dx \nonumber  \\
&=&  \frac{\gamma}{2\bar{F}^{\gamma-1}(t)}  J^w(X,t) + \frac{\gamma}{2\bar{F}^{\gamma+1}(t)}\int_t^\infty x f^2(x) \left[ 1- \bar{F}^{\gamma-1}(t) \right] dx.
\end{eqnarray}
The maxima of $J^w(X; t)$ exists, when $f(x) =\theta exp\{-x\theta \}$ and  $  \max\{  J^w(X; t)\}  =-\frac{1}{4}\left[  \theta t+\frac{1}{2} \right]. $ Thus, from Equation \eqref{max} the maxima of $J^w(X, Y; t)$ under PHR model also exists only when $f(x) = \theta exp\{-x \theta \}$, and $ \max \{ J^w(X, Y; t) \}=-\frac{\gamma }{2(\gamma+1)} \left[ \theta t+\frac{1}{\gamma+1}\right]. $  
\end{proof}
 %\hfill{$\Box$}

In the following, we express WDRJI in terms of the mean residual lifetime (MRL).
Let $X$ be a non-negative continuous random variable with survival function ${\bar{F}}$, such that $E(X) $ is finite. Then MRL of $X$ is defined {\color{red}in Equation \eqref{eq-mrl-j}} as
\begin{equation}\label{eq-mrl-j}
m(t)=\int^\infty_t \frac{\bar{F}(x)}{\bar{F}(t)}dx, \quad t \geq 0.
\end{equation}
The MRL function is of interest in many fields such as   survival analysis, actuarial studies, economic, reliability and so on. 
\begin{remark}\label{2}
\rm Let  $X$ and $Y$ be two non-negative random variables satisfying the PHR model. Then, we have
\begin{eqnarray}
J^w(X, Y; t)=-\frac{\gamma}{2}\int_t^\infty c^* x m(x)^{-2}dx,
\end{eqnarray}
where $c^*= \left[ 1+ m^{'}(x)  \right]^2 \left( \frac{\bar{F}(x)}{\bar{F}(t)} \right)^{\gamma+1}$.
% and $m(x)$ is mean residual lifetime.
\end{remark}

\begin{figure}[ht]
\centering
\includegraphics[scale=0.35]{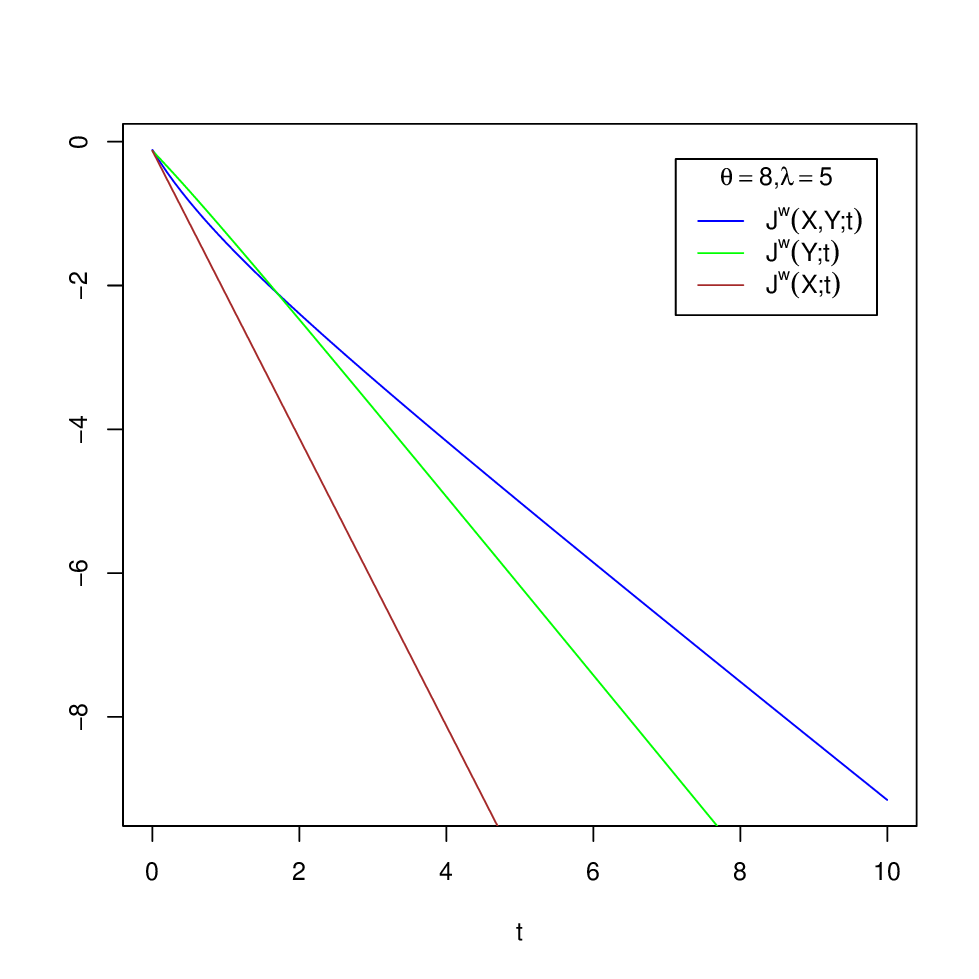}
\includegraphics[scale=0.35]{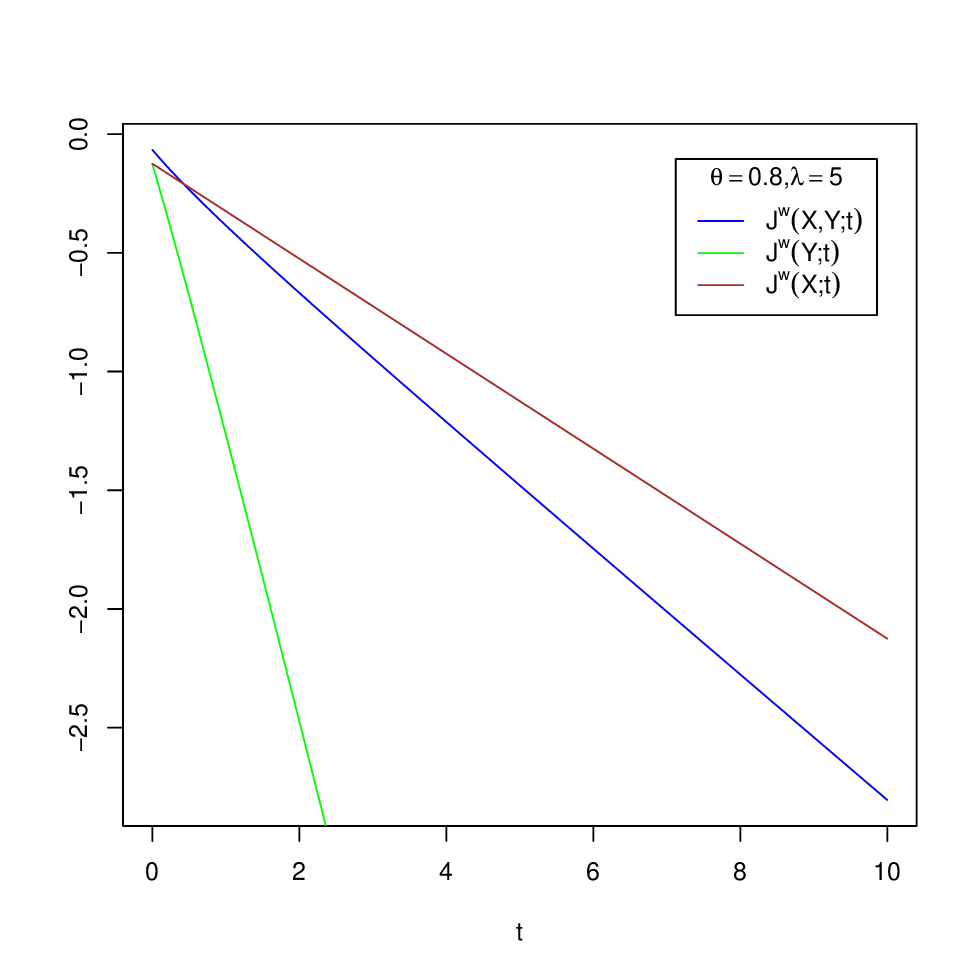}
\includegraphics[scale=0.35]{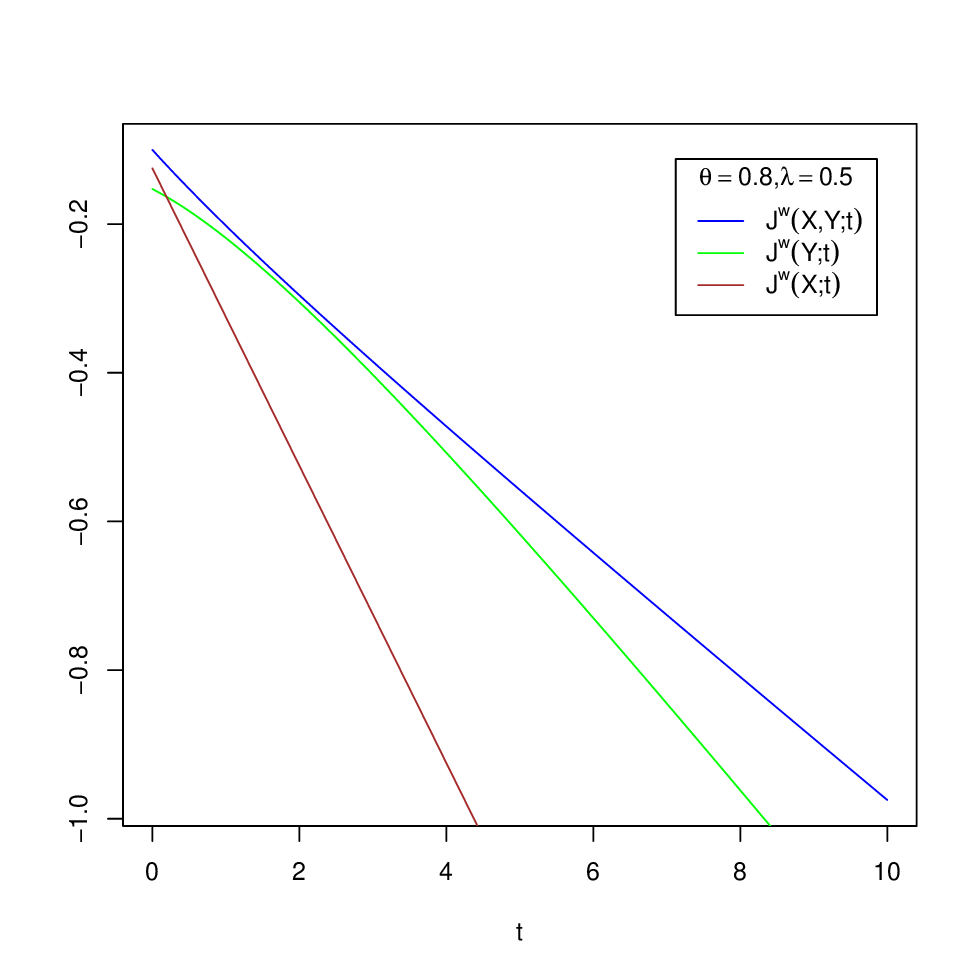}
\includegraphics[scale=0.35]{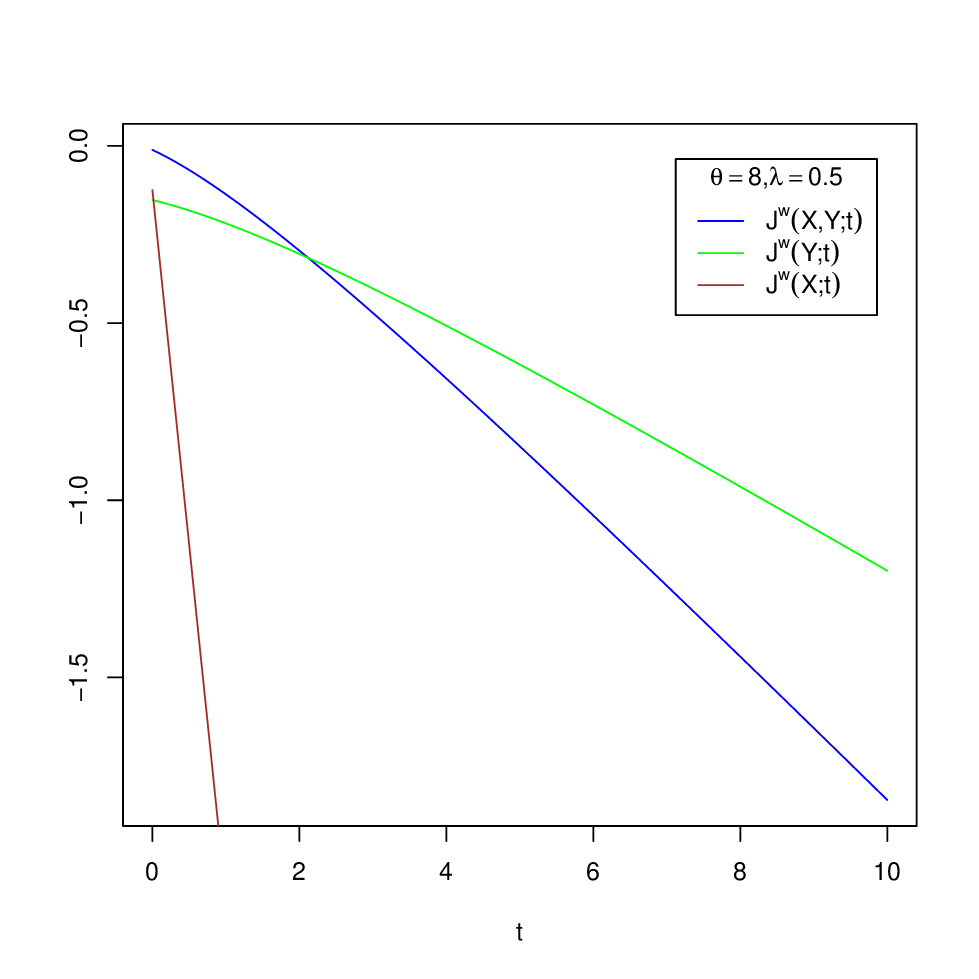}
\vspace{-0.5cm}
\caption{Graph of $ J^w(X, Y; t)$, $ J^w(X; t)$ and $ J^w(Y;t)$} for some selected values of $\lambda$ and $\theta$.\label{fig-ex-9}
\end{figure}

\example
\rm Let $X$ and $Y$ follow exponential and Lindley distributions, respectively, with SFs given by
\begin{align*}
&\bar{F}(t)=e^{-\theta t}; \quad  f(t)= \theta e^{-\theta t}, \ \ \theta >0, \ t\geq 0, \\ 
&\bar{G}(t)=(1+ \frac{\lambda}{\lambda+1}  t) e^{-\lambda t}; \quad
f(t)=\frac{\lambda^2}{\lambda+1}(1+t)e^{-\lambda t}, \ \  \lambda>0, \ t\geq 0.
\end{align*}
%where $\bar{\Gamma}_x(\alpha,\gamma)$ is known as the incomplete gamma function and defined as
After some algebraic manipulations, we have
\[J^w(X, Y; t)=-\dfrac{{\theta}{\lambda}^2\cdot\left(\left(t^2+t\right){\lambda}^2+\left(\left(2t^2+2t\right){\theta}+2t+1\right){\lambda}+\left(t^2+t\right){\theta}^2+\left(2t+1\right){\theta}+2\right)}{2\left({\lambda}+{\theta}\right)^3\left(\left(t+1\right){\lambda}+1\right)} ,\]
\[J^w(X, Y)= -\dfrac{{\theta}{\lambda}^2\cdot\left({\lambda}+{\theta}+2\right)}{2\left({\lambda}+1\right)\left({\lambda}^3+3{\theta}{\lambda}^2+3{\theta}^2{\lambda}+{\theta}^3\right)},\]
%\[\bar{J}(f|g; t)=.............  ,\]
 \[J^w(Y;t)=-\dfrac{\left(4t^3+8t^2+4t\right){\lambda}^3+\left(6t^2+8t+2\right){\lambda}^2+\left(6t+4\right){\lambda}+3}{16\left(\left(t+1\right){\lambda}+1\right)^2},\]
 \[J^w(X;t)=-\dfrac{2t{\theta}+1}{8}. \]
$J^w(X, Y; t)$, $J^w(Y,t)$ and $J^w(X,t)$ are shown in Figure \ref{fig-ex-9} for some selected values of  $\lambda$  and $\theta$.
{\color{red}Figure \ref{fig-ex-9}  shows that the inaccuracy measure WRJI and the residual extropy of both $X$ and $Y$ are decreasing in time $t$ and parameter $\theta$.}

\example
\rm Let a non-negative random variable $X$ be uniformly distributed over $(c, d)$, such that $c< d$, with  survival  and density  functions, respectively given by
$$ \quad  \bar{F}(x) =\frac{d-x}{d-c}, ~ x \in (c, d),$$  \quad
 and
 $$f (x) =\frac{1}{d-c}.$$
 If the random variables $X$  and $Y$ satisfy the PHR model, then the SF of the random variable $Y$ is
$$\bar{G}(x) = \bar{F}^\gamma(x) =\left[ \frac{d-x}{d-c} \right]^\gamma, ~ x \in (c, d), ~ \gamma \in (0,\infty).$$
Substituting these in Equation \eqref{sum1} and simplifying, we obtain WRJI measure as
\begin{eqnarray*}\label{90}
J^w(X,Y) &=& -\frac{\gamma}{2(d-c)^{\gamma+1}}\int_c^d x (d-x)^{\gamma-1} dx \\
&=&-\dfrac{\left(d-c\right)^{-{\delta}-1}\left(c{\delta}+d\right)\mathrm{e}^{\ln\left(d-c\right)\,{\delta}}}{2\left({\delta}+1\right)}\\
&=& -\frac{ (c \delta +d)\delta}{2(d-c)^\delta (\delta +1)}.
 \end{eqnarray*}

\begin{figure}[ht]
\centering
\includegraphics[scale=0.5]{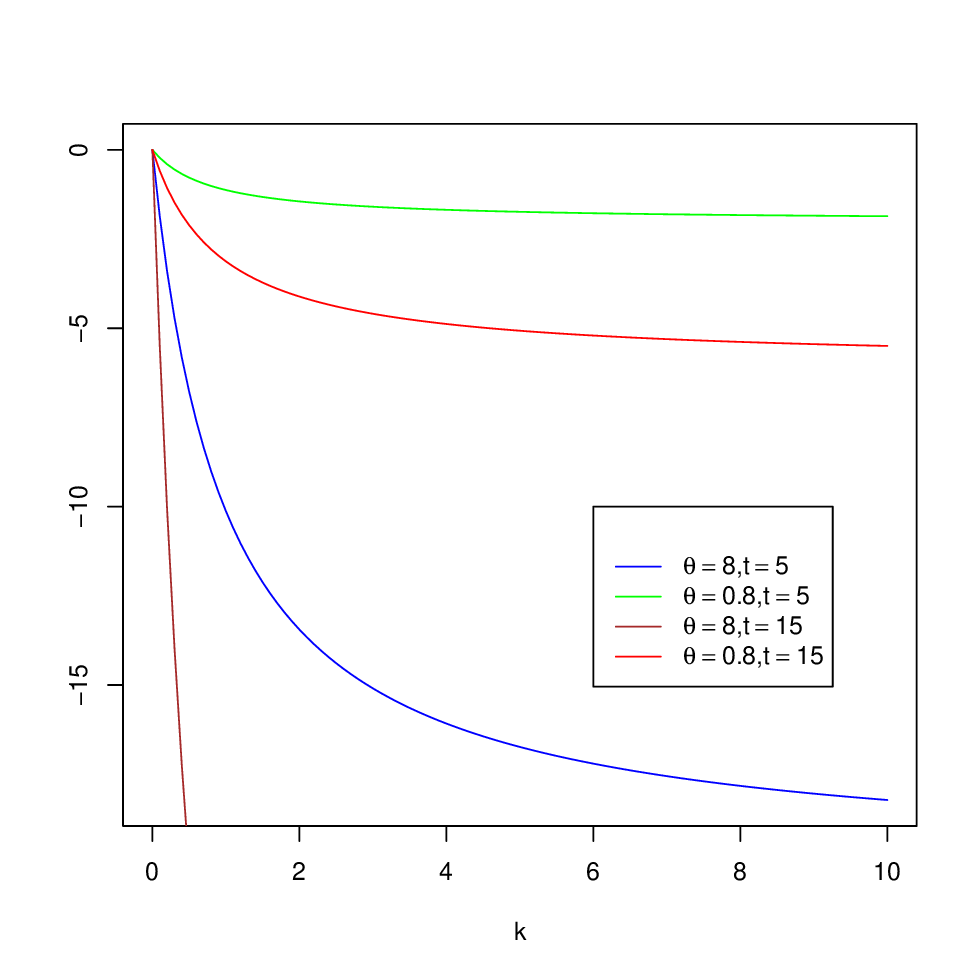}
%\vspace{-1cm}
\caption{Graph of $ J^w(X,Y; t)$ for some selected values of $\theta$ and $t$ as a function of $k$.}\label{fig-ex-11}
\end{figure}

In the sequel, we characterize uniform distribution in terms of the WRJI under the assumption that $X$ and $Y$ satisfy the PHR model.
Differentiating of Equation \eqref{H(f,g ; t)J} with respect to $t$ and using Equation \eqref{PRHRM},  we obtain
\begin{eqnarray}\label{uniform}
\frac{d}{dt}J^w(X,Y; t)
%&=&-\frac{\gamma}{2} \left[\frac{\int_t^\infty F^{\gamma+1}(x) \mu^2_F(x)dx}{F^{\gamma+1}(t)}\right] \nonumber \\
%&=& -\frac{\gamma}{2} \left[ - \mu^2_F(t) +(\gamma+1) \frac{f(t)}{F(t)} 
%  \frac{\int_t^\infty F^{\gamma+1}(x) \mu^2_F(x)dx}{F^{2(\gamma+1})(t)}   \right] \nonumber \\
%    &=& \frac{\gamma}{2}  \mu^2_F(t) +(\gamma+1) \mu^2_F(t)
%    \left[ - \frac{\gamma}{2} \int_t^\infty  \left( \frac{F(x)}{F(t)}  \right)^{\gamma +1}  \mu^2_F(x)dx   \right] \nonumber \\
= \frac{\gamma t}{2}  \lambda^2_F(t) +(\gamma +1) \lambda_F(t) J^w(X,Y; t).
\end{eqnarray}
This gives
$$\frac{d}{dt}J^w(X,Y; t)-\frac{\gamma }{2}t  \lambda^2_F(t) -(\gamma +1) \lambda_F(t) J^w(X,Y; t)=0.  $$
Hence for a fixed $t > 0$, $\lambda_F (t)$ is a solution of $g^w(x)=0$, where
\begin{eqnarray}\label{uni2}
g^w(x) = \frac{d}{dt}J^w(X,Y; t)-\frac{\gamma}{2} x^2 t-(\gamma +1) x J^w(X,Y; t). 
\end{eqnarray}
Differentiating both side of Equation \eqref{uni2} with respect to $x$, we get
\begin{eqnarray}\label{g*}
\frac{d}{dx} g^w(x) =- \gamma t x -\left( \gamma +1\right) J^w(X,Y; t).
\end{eqnarray}
Thus $\frac{d}{dx} g^w(x) =0$ gives, 
%$x=-\frac{\gamma+1}{\gamma}\bar{J}(f,g; t)=x^*$.
\begin{eqnarray}
x=-\frac{\gamma+1}{\gamma t}  J^w(X,Y; t)=x_0.
\end{eqnarray}
In the following, we give a theorem which characterizes uniform distribution in terms of WRJI.
\begin{theorem}
\rm Suppose  non-negative continuous random variables $X$ and $Y$ satisfy the PHR model in Equation \eqref{PRHRM}. Then random variable $X$ over $(c, d)$ such that $c< d$ has uniform distribution if and only if 
\begin{eqnarray}\label{thm3}
J^w(X,Y;t)=\dfrac{{\gamma}t+d}{2\left({\gamma}+1\right)\left(t-d\right)}.
\end{eqnarray}
\end{theorem}
\proof
The only if part of the theorem is straightforward since in the case of uniform distribution of random variable $X$ over $(c, d)$
$$  f(x)= \frac{1}{d-c}, \quad  \bar{F}(x)=\frac{d-x}{d-c}.  $$
Hence, under PHR model, $G(x)=\left( \frac{d-x}{d-c}\right)^\gamma$ which gives $ g(x)=\frac{\gamma }{(d-c)^\gamma} (d-x)^{\gamma-1.}   $
Substituting these in Equation \eqref{H(f,g ; t)J} and simplifying, we get
\begin{eqnarray}\label{th-unif}
J^w(X,Y;t)=\dfrac{{\gamma}t+d}{2\left({\gamma}+1\right)\left(t-d\right)}.
\end{eqnarray}
To prove the if part let Equation \eqref{uni2} be valid. Then from Equation \eqref{thm3}, we have 
$g^w(0) =\frac{d}{dt}J^w(X,Y; t) < 0. $ 
Also we can show that $g^w(x)$ is a  concave function with maximum occurring at $x =x_0$.  Thus  $g^w(x)=0$ has unique solution if $g^w(x_0)=0$. We have $x_0=- \left(  \frac{\gamma+1}{\gamma} \right) \frac{J^w(X,Y; t)}{t}.$ Using Equation \eqref{th-unif}, we get $x_0=\left( d-t \right)^{-1},~ t<d$  and   
\begin{eqnarray}\label{uni3}
g^w(x) = \frac{d}{dt} J^w(X,Y; t)-\frac{\gamma}{2}t x{^2}-(\gamma +1) x J^w(X,Y; t) =0.
\end{eqnarray}
Thus, $g^w(x)=0$ has unique solution given by $x=x_0$. But $\lambda_F(t)$ is a solution to Equation \eqref{uni2}. Hence $\lambda_F(t) = x_0=1/(d-t),~ t <d$ is the unique solution to $g^w(x)=0$. So, the distribution is uniform. %\hfill{$\Box$}
% and this proves the result. 
To illustrate  the characterization results obtained above, we consider the following example.
\example \label{3.1}
\rm Let  random variables $X_1, X_2,\ldots,X_k$ have exponentially distributed  with  PDF $f (x)=\theta \exp\{-\theta  x\}$ and 
CDF $F(x)= 1-\exp\{-\theta x\}, ~ x>0, \theta >0$ representing the lifetime of components, in a $k$-component series system, then the lifetime of the system is given by $Y=\min(X_1, X_2, \cdots , X_k)$.
% with CDF,  $G(x)$ given by \eqref{PRHRM}. 
 If $G$ is the CDF for $Y,$ then under PHR model the CDF of $Y$ and its PDF are given  $\bar{G}(x)=\bar{F}^k(x)$ and  $g(x) = k f (x) \bar{F}^{k-1}(x)$, respectively.
 % \begin{align*}
%&G(x)=[F(x)]^n; \quad g(x) = m f (x) [F(x)]^{m-1}, \     t\geq 0, \ \theta>0, \\
%%& \bar{G} (t)=e^{-\lambda t}; \quad f(t)= \lambda e^{-\lambda t}, \   t\geq 0, \ \lambda>0.
%\end{align*}
 %\begin{align*}
%& \bar{F} (t)=e^{-\theta t}; \quad f(t)= \theta e^{-\theta t}, \     t\geq 0, \ \theta>0, \\
%& \bar{G} (t)=e^{-\lambda t}; \quad f(t)= \lambda e^{-\lambda t}, \   t\geq 0, \ \lambda>0.
%\end{align*}
From Equation \eqref{H(f,g ; t)J} under PHR model, we given
%\begin{align*}
%  J(f,g)= -\frac{ \lambda \theta }{2(\theta+\lambda)},
%\end{align*}
%and
\begin{eqnarray}
%\bar{J}(f,g)= 
%-\dfrac{{\theta}\cdot\left(\mathrm{e}^{{\theta}x}+m\right)\mathrm
%{e}^{m\ln\left(\left|\mathrm{e}^{-{\theta}x}-1\right|\right)-{\theta}x}}{2\left(m+1\right)}\\
J^w(X,Y;t)= -\frac{k \theta^2}{2e^{-t \theta(k+1)} }
\int_t^\infty x e^{-x \theta(k+1)}dx
= -\dfrac{k\left(\left(k+1\right)t{\theta}+1\right)}{2\left(k^2+2k+1\right)}.     
\end{eqnarray}
Taking $limit$ as $k \rightarrow \infty $, we obtain
\begin{equation}
\lim_{k \rightarrow \infty} J^w(X,Y;t) =-\dfrac{t\theta}{2}.
\end{equation}
{\color{red}Figure \ref{fig-ex-11}  shows that when $k$, the number of components increases in a series system then the magnitude of the WRJI decreases. Also, it is observed that the inaccuracy measure WRJI is decreasing in time $t$ and parameter $\theta$.}

\section{Some bounds and inequalities for WRJI}
In this section, upper and lower bounds and some inequalities concerning weighted dynamic residual inaccuracy measure are determined.
%In the sequel, we express some upper(lower) bounds for WDRJI in terms of the extropy.
In the sequel, we express some lower bounds for WRJI in terms of the HRF.
\begin{corollary}\label{22}
Suppose  $X$ and $Y$ are two non-negative random variables satisfying the PHR model. Then, we have
\begin{eqnarray}
J^w(X, Y; t)   \geq  -\frac{\gamma}{2} \int_t^\infty x   \lambda_F^2(x)   dx.
\end{eqnarray}
\end{corollary}
\begin{proof}
We known that $t<x$ then  $F(t)<F(x)$. This implies $\bar{F}(x)/ \bar{F}(t) <1$. 
Thus,   $[\bar{F}(x)/ \bar{F}(t)] ^{\gamma+1}<\bar{F}(x)/ \bar{F}(t)$. After some  calculations and using Equation \eqref{J(f, g; t)phr}, the proof is completed.
 \end{proof}
\begin{remark}\label{22}
\rm Suppose  $X$ and $Y$ are two non-negative random variables satisfying the PHRM. Then, we given
\begin{eqnarray}
J^w(X, Y; t)   \geq  -\frac{\gamma}{2} \int_t^\infty x  \left[ -\log \bar{F}(x)\right]^2  dx.
\end{eqnarray}
\end{remark}
\begin{proposition}\label{4}
\rm Let $X$ and $Y$ be two non-negative random variables. Then, we have
\begin{eqnarray}
J^w(X, Y; t)   \geq  \left[\bar{F}(t) \bar{G}(t)  \right] J^w(X, Y).
\end{eqnarray}
\end{proposition}

\begin{figure}[ht]
\centering
\includegraphics[scale=0.5]{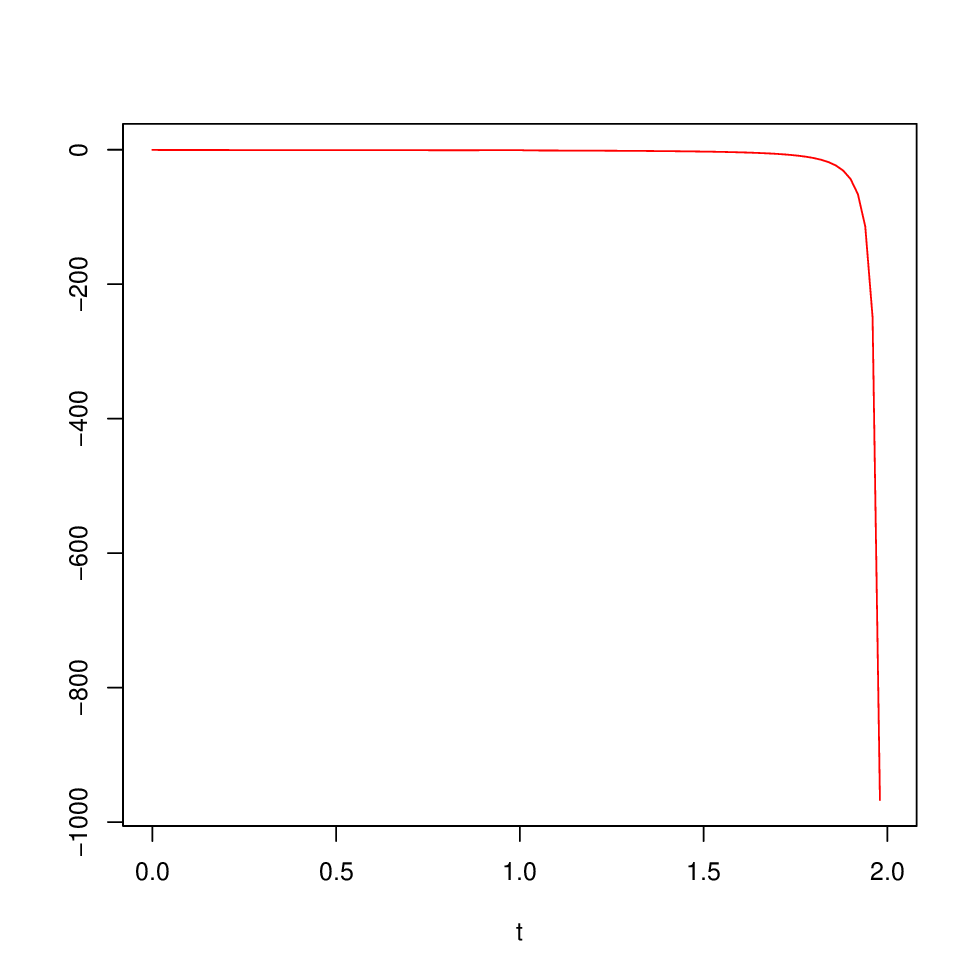}
\vspace{-0.5cm}
\caption{Graph of $J^w(X,Y; t)$ in Example \ref{3.2}.}\label{fig-ex-12}
\end{figure}

In the following, we consider another example where $F(x)$ and $G(x)$ does not satisfy PHR model.
\example \label{3.2}
\rm Suppose $X$ and $Y$ are two non-negative random variables having distribution functions, respectively
%\[
% F_X(x)=   \left \{
% \begin{array}{ll}
% \frac{x^2}{2}     &   \quad   \text{,~ $ 0 \leq x<1 $} \\
% \frac{x^2+2}{6}     &     \quad   \text{,~     $ 1 \leq x<2 $    }\\
% 1      &     \quad   \text{,~     $ x \geq 2 $   },
% \end{array}  \right. 
%   \]
%and
%\[
% G_Y(x)=   \left \{
% \begin{array}{ll}
% \frac{x^2+x}{4}     &   \quad   \text{,~ $ 0 \leq x<1 $} \\
%\frac{x}{2}     &     \quad   \text{,~      $ 1 \leq x<2 $   }\\
% 1    &     \quad   \text{,~     $ x \geq 2 $      },
% \end{array}  \right. 
%   \]
%The DPJI and PJI measures are given by
%\[
% J(f, g; t) = \left \{
% \begin{array}{ll}
% -................    &   \quad   \text{,~ $ 0 \leq x<1 $} \\
% 
%-..............      &     \quad   \text{,~      $ 1 \leq x<2 $   }\\
%0 &     \quad   \text{,~     $ x \geq 2 $      },
% \end{array}  \right. 
%   \]
%and
%\[
% J(f, g)=\left \{
% \begin{array}{ll}
% -............   &   \quad   \text{,~ $ 0 \leq x<1 $} \\
% -...............   &     \quad   \text{,~      $ 1 \leq x<2 $   }\\
% 0  &     \quad   \text{,~     $ x \geq 2 $      }.
% \end{array}  \right. 
%   \]
%
%
%%%%%%%%%%%%%%%%%%%%%%%%%
%
%Or : let
%        \textcolor{blue}{   $1<x<t$ ...................................}
%\textcolor{blue}{}
\[
 F_X(x)=   \left \{
 \begin{array}{ll}
 \frac{x^2}{2}     &   \quad   \text{,~ $ 0 \leq x<1 $} \\
 \frac{x^2+2}{6}     &     \quad   \text{,~     $ 1 \leq x<2 $    }\\
 1      &     \quad   \text{,~     $ x \geq 2 $   },
 \end{array}  \right. 
   \]
and
\[
 G_Y(x)=   \left \{
 \begin{array}{ll}
 \frac{x^2+x}{4}     &   \quad   \text{,~ $ 0 \leq x<1 $} \\
\frac{x}{2}     &     \quad   \text{,~      $ 1 \leq x<2 $   }\\
 1    &     \quad   \text{,~     $ x \geq 2 $      },
 \end{array}  \right. 
   \]
The WDRJI and WRJI measures are given by
\[
 J^w(X, Y; t) = \left \{
 \begin{array}{ll}
 -\dfrac{4-4(4t^{3}-3t^{2})}{3(2-t^{2})(4-t^{2}-t)}-\dfrac{3}{2(4-t^{2})(2-t)}   &   \quad   \text{,~ $ 0 \leq t<1 $} \\
 
-\dfrac{4-t^{2}}{2(4-t^{2})(2-t)}  &     \quad   \text{,~      $ 1 \leq t<2 $   }\\
0 &     \quad   \text{,~     $ t \geq 2 $      },
 \end{array}  \right. 
   \]
and
\[
 J^w(X, Y)=\left \{
 \begin{array}{ll}
 -\frac{17}{48}   &   \quad   \text{,~ $ 0 \leq x<1 $} \\
 -\frac{1}{4}   &     \quad   \text{,~      $ 1 \leq x<2 $   }\\
 0   &     \quad   \text{,~     $ x \geq 2 $      }.
 \end{array}  \right. 
   \]
Figure \ref{fig-ex-12} provides the graphs of $J^w(X, Y; t)$ as a function of $t$. Notice $J^w(X,Y;t)$ is a decreasing and continuous function in terms of $t$.$\hfill \Box$

In the sequel,  we observe the following relation between three inaccuracy measures considered in this paper.
\begin{remark}
\rm Let $X$ and $Y$ be two non-negative continuous random variables with PDFs respectively $f(x)$ and $g(x)$. Suppose $F(x) $ and $G(x)$ are their CDFs, respectively. The relation between the three inaccuracy measures  given by
\begin{eqnarray}
J^w(X,Y)= F(t)G(t) \bar{J}^w(X,Y;t) + \bar{F}(t)\bar{G}(t) \bar{J^w}(X,Y;t).
\end{eqnarray}
\proof
From Equation \eqref{sum1}, we have
\begin{eqnarray}
J^w(X,Y)&=&-\frac{1}{2}\int_0^t x f(x)g(x)dx -\frac{1}{2}\int_t^\infty x f(x)g(x)dx \nonumber \\
&=&-\frac{1}{2}   F(t)G(t) \int_0^t  x\frac{f(x)g(x)}{F(t)G(t) } dx
-\frac{1}{2} \bar{F}(t)\bar{G}(t) \int_t^\infty  x \frac{f(x)g(x)}{\bar{F}(t)\bar{G}(t) } dx \nonumber \\
&=& F(t)G(t) \bar{J^w}(X,Y,t)+\bar{F}(t)\bar{G}(t){J^w}(X,Y,t), \nonumber  
\end{eqnarray}
where $\bar{J}^w(X,Y,t)$ is weighted dynamic past inaccuracy measure.
\end{remark}
%In the sequel, we express  for DRJI in terms of the  measure of inaccuracy equation \eqref{sum1}.

\rm In the next remarks, the relationship between $J^w(X, Y; t)$ and $J^w(X,Y)$ are presented.
\begin{remark}\label{26}
\rm Let $X$ and $Y$ be two non-negative random variables satisfying the PHR model. Then, we have
\begin{eqnarray}
J^w(X, Y; t) = c_3(t) J^w(X, Y) + c_4(t), 
\end{eqnarray}
where $ c_3(t)=\bar{F}^{-(\gamma+1)}(t) $ and $c_4=\frac{\gamma}{2}\int_0^t x \frac{f^2(x)\bar{F}^{\gamma-1}(x)}{\bar{F}^{\gamma+1}(t)}dx$.
\end{remark}

\begin{remark}\label{27}
\rm Suppose $X$ and $Y$ are two non-negative random variables satisfying the PHR model. Then, we given
\begin{eqnarray}
J^w(X,Y;t) = b_1(t) [  J^w(X,Y) + b_2(t)], 
\end{eqnarray}
where $ b_1(t)=\bar{F}^{-(\gamma+1)}(t) $ and $b_2=\frac{\gamma}{2}\int_0^t  x \mu^2(x) \bar{F}^{\gamma+1}   dx.$
\end{remark}

In order to provide a lower bound for the WRJI measure of a random variable $X$, we study the following conditional mean value (Vitality function)
\begin{equation}\label{vitality}
V(X;t):=E(X \mid X>t)=\frac{1}{\bar{F}(t)} \int_t^{\infty} x f(x) d x,
\end{equation}
a result which finds applications in insurance and economics. For more details, refer to {\color{red}Furman and Zitikis} \cite{Furman}.

\begin{theorem}
\rm If the hazard rate function $\lambda_G(x)$ is decreasing in $x$, then
$$
J^{w}(X, Y ; t) \geq-\frac{1}{2}\ \lambda_G(x) V(X;t).
$$
\end{theorem}
\begin{proof}
 $J^{w}(X,Y;t)$ can be rewritten as
$$
J^w(X, Y ; t)=-\frac{1}{2} \int_t^{\infty} x \dfrac{f(x)}{\bar{F}(t)} \lambda_{G}(x) \dfrac{\bar{G}(x)}{\bar{G}(t)} dx
$$
Since $\frac{\bar{G}(x)}{\bar{G}(t)} \leq 1$, for $x \geq t$, and also by the assumption that $\lambda_G(x)$ is a decreasing function, the proof is completed.
\end{proof}

\example
\rm If true distribution function $F(x)$ and reference distribution function $G(x)$ are exponentially distributed with parameters $\lambda_1>0$ and $\lambda_2>$ 0 respectively, then the WRJI measure is derived as follows.
%$$
%\begin{aligned}
%f(x)  =\lambda_1 e^{-\lambda_1 x}, \quad g(x)=\lambda_2 e^{-\lambda_2 x}, \\
%F(x) =1-e^{-\lambda_1 x}, 
%G(x) =1-e^{-\lambda_2 x} .
%\end{aligned}
%$$
%Substituting for $\bar{G}(x), \bar{F}(x), f(x)$, and $g(x)$ in \eqref{H(f,g ; t)J}, 
we obtain
\begin{equation}\label{vitality}
J^w(X, Y ; t)=-\dfrac{{\lambda}_1{\lambda}_2\cdot\left(\left({\lambda}_2+
{\lambda}_1\right)t+1\right)}{2\left({\lambda}_2+{\lambda}_1\right)^2}.
\end{equation}
% Therefore equation \eqref{vitality} holds.
Note that the hazard rate function is constant for an exponential distribution, that is, $\lambda(t)=\lambda$, and the conditional mean value $E(X \mid X>t)= t+\frac{1}{\lambda}$.

%\section{Bounds and Inequalities }
In this part, we obtain some lower and upper bounds for measure of inaccuracy  between $X$ and $Y$. First, we express an upper (a lower) bound for WRJI in terms of the extropy.
\begin{theorem}\label{p2}
Let $X$ and $Y$ be two non-negative random variables satisfying the PHR model. Then, we have
%\begin{eqnarray}
\begin{itemize}
\item[(i)] For $\gamma > 1$,~ $J^w(X, Y; t) \geq \gamma   J^w(X; t)$, 
\item[(ii)]  For $ 0< \gamma \leq 1$,~ $J^w(X, Y; t) \leq  \gamma J^w(X; t)$.
\end{itemize}
%\end{eqnarray}
\end{theorem}
\begin{proof}
Since $t<x$ then $\bar{F}(x)<\bar{F}(t) $. Also,  $\bar{F}(x)/ \bar{F}(t) <1$. 
Therefore, for  $ 0< \gamma \leq 1$,  $[\bar{F}(x)/ \bar{F}(t)] ^{\gamma-1}>1$  and for   $\gamma > 1$, $[\bar{F}(x)/ \bar{F}(t)] ^{\gamma-1}<1$. 
 After some algebraic manipulations and using Definition $4$,  the proof is completed.
 \end{proof}
 
In the following, we express a lower bound for $J^w(X,Y; t)$ in terms of the weighted extropy inaccuracy.
\begin{remark}
\rm A lower bound for the  WDRIJ  between the distributions $X$ and $Y$   is obtained
\begin{eqnarray}
J^w(X,Y;t)  \geq   a(t)  J^w(X,Y), 
\end{eqnarray}
where $a(t)=[ \bar{F}(t) \bar{G}(t)]^{-1}.$
\end{remark}
\begin{proof}
\begin{eqnarray*}
J^w(X,Y; t)  &=& - \frac{1}{2} \int^\infty_t x  \frac{f(x)}{\bar{F}(t)}  \frac{g(x)}{\bar{G}(t)} dx \\
&\geq &
 - \frac{1}{2} \int^\infty_0 x  \frac{f(x)}{\bar{F}(t)}  \frac{g(x)}{\bar{G}(t)} dx \\
&=&[\bar{F}(t) \bar{G}(t)]^{-1} J^w(X,Y). 
\end{eqnarray*}
\end{proof}

In the following, we express some lower bounds for WRJI in terms of the HRF.
%In the sequel, we express an upper(lower) bound for DRJI in terms of the extropy.
\begin{corollary}\label{22}
\rm Let that $X$ and $Y$ be two non-negative random variables satisfying the PHR model. Then, we have
\begin{eqnarray}
J^w(X, Y; t)   \geq   c_1 \int_t^\infty   \bar{F}(x) \mu_F^2(x)   dx,
\end{eqnarray}
where $c_1=-\frac{\gamma}{2\bar{F}(t)} $
\end{corollary}
\begin{proof}
We know that for $t<x$ then $F(t)<F(x)$. This implies  $\bar{F}(x)/ \bar{F}(t) <1$. 
Thus,   $[\bar{F}(x)/ \bar{F}(x)] ^{\gamma+1}<\bar{F}(x)/ \bar{F}(x)$. After some calculations and using {\color{red}Equation} \eqref{J(f, g; t)phr}, the proof is completed.
\end{proof}

\begin{proposition}\label{23}
\rm Let $X$ and $Y$ be two non-negative random variables satisfying the PHR model. Then, we have
\begin{eqnarray}
J^w(X,Y;t)   \geq   -\frac{\gamma}{2} \int_t^\infty x \mu_F^2(x)   dx.
\end{eqnarray}
\end{proposition}
\begin{proof}
 We know that $ 0 \leq \bar{F}(x) \leq 1  $ and since $t<x$ then   $[\bar{F}(x)/ \bar{F}(t)] ^{\gamma+1}<1$.  After some algebraic manipulations and using Definition $4$,  the proof is completed.
 \end{proof}

\begin{remark}\label{29}
\rm Let $X$ and $Y$ be two non-negative random variables satisfying the PHR model and decreasing PDFs such that $f (0) \leq1$. Then, we have
\begin{eqnarray}
J^w(X, Y; t) \geq c_2 \int_t^\infty x \bar{F}^{\gamma-1}(x)dx,
\end{eqnarray}
where $ c_2=-\frac{\gamma}{2\bar{F}^{\gamma+1}(t) }$.
\end{remark}

\begin{remark}
\rm  Let $X$ and $Y$ be two non-negative continuous random variables with PDFs  $f$ and $g$, respectively. Then, we have
%\begin{eqnarray}
\begin{itemize}
\item[(i)]  $J^w(X, Y; t) \geq  J^w(X; t)$, 
\item[(ii)]   $J^w(X| Y; t) \leq  - J^w(X; t)$.
\end{itemize}
\end{remark}

\begin{proposition}\label{24}
\rm Let $X$ and $Y$ be two non-negative random variables satisfying the PHR model. Then, we have
\begin{eqnarray}
J^w(X, Y; t)   \geq   d_1 \int_t^\infty x \mu_F(x) f(x)   dx,
\end{eqnarray}
where $d_1=-\frac{\gamma}{2\bar{F}(t)} $
\end{proposition}
%\begin{proof}
% We known that $ 0 \leq \bar{F}(x) \leq 1  $ and because $t<x$ then   $[\bar{F}(x)/ \bar{F}(x)] ^{\gamma+1}<1$.  After some algebraic manipulations and using definition \eqref{111},  the proof is completed.
% \end{proof}
%In the following, we express an lower bound for DRJI in terms of the extropy.
\begin{proposition}\label{p4}
\rm Suppose $X$ and $Y$ are two non-negative random variables satisfying the PHR model. We have 
\begin{itemize}
\item[(i)] $J^w(X, Y; t) \geq d_2\int_t^\infty x \mu^2_F (x) dx$,
\item[(ii)] $ J^w(X, Y; t) \geq d_2\int_t^\infty x \mu_F (x) f(x)dx$,
\end{itemize}
where $d_2=-\frac{\gamma}{2\bar{F}(t)}$
\end{proposition}
%\begin{proof}
%The proof is simply done as
%\end{proof}
In the sequel, we express a lower bound for WRJI in terms of the MRL.
\begin{theorem}\label{25}
\rm Let $X$ and $Y$ be two non negative random variables satisfying the PHR model and decreasing PDFs  such that $f (0) \leq1$. For $\gamma=2$, we given  
\begin{eqnarray}
J^w(X, Y; t) \geq -k_1 m(t),
\end{eqnarray}
where $k_1=\bar{F}^{-2}(x)$.
\end{theorem}

In the following, we express an upper bound for WRJI in terms of the  measure of inaccuracy  {\color{red}in Equation} \eqref{sum1}.

\begin{proposition}\label{40}
\rm Let $X$ and $Y$ be non-negative continuous random variables satisfying the PHR model. Then, we have
\begin{eqnarray}
J^w(X, Y; t) \leq k_2 J^w(X, Y),
\end{eqnarray}
where $k_2=\frac{1}{\bar{F}^{(\gamma+1)}(t)}$.
\end{proposition}

%
%In sequel, we provide two lower bounds for DRJI of non negative random variables in terms of RJI and MRL functions.????????

%--------------
%\begin{equation}\label{1000}
% 0< \gamma \leq 1$,  $[\bar{F}(x)/ \bar{F}(x)] ^{\gamma-1}>1
%\end{equation}
% and for
% \begin{equation}\label{1001}
%  \gamma > 1$, $[\bar{F}(x)/ \bar{F}(x)] ^{\gamma-1}<1. 
% \end{equation}
% 
%  Multiplying both sides of \eqref{} by −t/2 and integrating with respect to t from $t$ to $\infty$, then we have
% ---------------------------------------

\section{Non-parametric estimator}\label{sec-non-para}
We defined WRJI in Equation \eqref{H(f,g ; t)J}. In this section, we consider estimating of this parameter.
The problem of estimation of $f(x)$ is more complicated than that of $\bar{F}(x)$. For this case, a method known as kernel density estimation is used.
Let $(X_{1},X_{2},...,X_{n})$ be a random sample with pdf $f(x)$ and SF $\bar{F}(x)$. Then, an estimator for $f(x)$ can be given as
$$
f_{n}(x)=\dfrac{1}{nh_n}\sum_{i=1}^{n}K(\dfrac{x-X_{i}}{h_n}),
$$
where $h_n$ is a bandwidth satisfied the condition that $h_n$ tends to 0 as $n$ goes to infinity, and $K(x)$ is a kernel function satisfied the following conditions
\begin{align*}
\int_{-\infty}^{\infty} |K(x)|dx<\infty,  \\
\sup_{-\infty < x < \infty}|K(x)| < \infty, \\
\lim_{x \rightarrow \infty}|xK(x)|=0.
\end{align*}
For additional details on this concept, readers can refer to {\color{red}Rosenblatt} \cite{rosen-56} and {\color{red}Parzen} \cite{parzen-62}.
We consider two estimators for CDF ${F}(x)$.
One can use an empirical distribution function (ECDF) for estimating $\bar{F}(x)$. Then, ECDF for estimating $\bar{F}(x)$ can be computed as 
$$
\bar{F}_{n}(t)=\dfrac{\sum_{i=1}^{n}I(X_{i}>t)}{n},
$$
where $I(x)$ is an indicator function takes 1 for non-negative $x$ and 0 for otherwise.
As $\bar{F}_{n}(t)$ is a step function, some researchers considered a smoothed version of ECDF based on the kernel estimation method. Let $K(.)$ be a kernel density function. Then an estimation for $F(.)$ can be obtained as
\begin{equation}
\hat{F}_{h}(x)=\frac{1}{n}\sum_{i=1}^{n} W(\frac{x-X_{i}}{h_{n}}),
\end{equation}
where $h_n$ is a bandwidth parameter and $W(x)$ is defined as
$$
W(x)=\int_{-\infty}^{x}  K(t) dt.
$$
Therefore, we consider two estimators for the proposed measure WRJI as
\begin{equation}\label{eq-est-ecdf}
J^{w}_{n}(X,Y;t) =-\frac{1}{2} \int^\infty_t  x  \frac{f_{n}(x)}{\bar{F}_{n}(t)}\frac{g_{n}(x)}{\bar{G}_{n}(t)} dx,
\end{equation}
and 
\begin{equation}\label{eq-est-ecdf}
J^{w}_{h}(X,Y;t) =-\frac{1}{2} \int^\infty_t  x  \frac{f_{n}(x)}{\hat{\bar{F}}_{h}(t)}\frac{g_{n}(x)}{\hat{\bar{G}}_{h}(t)} dx.
\end{equation}
The choice of kernel function is not crucial for estimating $f(.)$ and $F(.)$, but it is the case for bandwidth $h_n$. In this paper, we apply the cross validation method to find the best $h_n$ for both $f(.)$ and $F(.)$.
In the case of $f(.)$, the best $h_n$ can be obtained by minimizing the mean integrated squared error as
\begin{equation} \label{eq-hf-pdf}
h_{f} = \arg \min _{h_n} E\left[ \int f_{n}^2 (x) dx - \frac{2}{n} \sum_{i=1}^{n} f_{n,-i} (X_{i}) \right],
\end{equation}
where $f_{n,-i}(X_{i})$ is the kernel estimation obtained by omitting $X_{i}$. Indeed the $h_f$ in Equation \eqref{eq-hf-pdf} is the optimal bandwidth for non-normal data and as we see the cross validation method for finding the best bandwidth is a data driven approach.
Similar argument can be applied for finding the best bandwidth when we need to estimate the CDF $F(.)$. In this case, $h_F$ can be obtained as
\begin{equation} \label{eq-hF-cdf}
h_{F} = \arg \min _{h_n} \frac{1}{n} \sum_{i=1}^{n} \int \left(I(x-X_{i} \geq 0)- \hat{F}_{h,-i}(x) \right)^2 dx,
\end{equation}
where $F_{h,-i}(x)$ is the kernel estimation of $F$ obtained by omitting $X_{i}$. One can refer to {\color{red}Bowman et al.} \cite{bow-hal-prv-98}, for additional details and discussions on this subject.
% % % % % % % % % % % % % % % % %
\subsection{Simulation}
We run a simulation with $10000$ iterations to compute $J^{w}_{n}(X,Y;t)$ and $J^{w}_{h}(X,Y;t)$. We consider some distributions such as exponential (exp$(\lambda)$) and beta (beta$(\alpha,\beta)$) with sample size $n$ varies in $\{30,50\}$, and then the bias and the mean squared error (MSE) of estimates are provided. The Gaussian kernel function is used  and the bandwidths $h_n$ are considered via the method of cross validation as proposed in Section \ref{sec-non-para} for estimating of PDF $f(.)$ and CDF $F(.)$, respectively, based on two estimators $J^{w}_{n}(X,Y;t)$ and $J^{w}_{h}(X,Y;t)$. The results of simulation are provided in Table \ref{tab-1}. For different values of time $t$ given in Table \ref{tab-1}, in the case of exponential, we compare the proposed estimators $J^{w}_{n}$ and $J^{w}_{h}$, when we consider an exp$(\lambda=1)$ as the actual distribution and exp$(\lambda)$ for $\lambda=2,5,7$ as the one that assigned by an experimenter. Similar comparison is provided for the case of beta distribution in Table \ref{tab-1}. In this case, a beta$(1,1)$ is considered as the actual distribution.

\begin{table}[h]
\centering
\caption{Bias and MSE for $J^{w}_{n}(X,Y;t)$ and $J^{w}_{h}(X,Y;t)$ for exponential and beta distributions.}\label{tab-1}
\begin{tabular}{cccccccccccc}\hline
\multicolumn{3}{c}{exponential} &  & \multicolumn{2}{c}{$\lambda=2$} &  & \multicolumn{2}{c}{$\lambda=5$}&  & \multicolumn{2}{c}{$\lambda=7$}   \\ \cline{1-3} \cline{5-6} \cline{8-9} \cline{11-12} %\hline
 $t$  &  $n$ &  &  & $J^{w}_{n}$ & $J^{w}_{h}$ &  & $J^{w}_{n}$ & $J^{w}_{h}$ &  & $J^{w}_{n}$ & $J^{w}_{h}$    \\ \hline
0.01&  30& bias &  &0.0513  &0.0404  &  &0.0551  &0.0518  &  &0.0456   &0.0432    \\
    &    & MSE  &  &0.0027  &0.0018  &  &0.0035  &0.0027  &  &0.0021   &0.0019    \\
    &  50& bias &  &0.0490  &0.0386  &  &0.0543  &0.0512  &  &0.0452   &0.0431    \\ 
    &    & MSE  &  &0.0025  &0.0016  &  &0.0030  &0.0026  &  &0.0021   &0.0019    \\ 
    &    &  &  &  &  &  &  &  &  &   &    \\
0.05&  30& bias &  &0.0561  &0.0456  &  &0.0671  &0.0637  &  &0.0588   &0.0562    \\
    &    & MSE  &  &0.0033  &0.0023  &  &0.0045  &0.0041  &  &0.0035   &0.0032    \\
    &  50& bias &  &0.0537  &0.0439  &  &0.0665  &0.0634  &  &0.0581   &0.0557    \\ 
    &    & MSE  &  &0.0030  &0.0021  &  &0.0044  &0.0040  &  &0.0034   &0.0031    \\ 
    &    &  &  &  &  &  &  &  &  &   &    \\ 
0.10&  30& bias &  &0.0618  &0.0512  &  &0.0820  &0.0787  &  &0.0744   &0.0716    \\
    &    & MSE  &  &0.0047  &0.0029  &  &0.0068  &0.0062  &  &0.0056   &0.0052    \\
    &  50& bias &  &0.0586  &0.0499  &  &0.0808  &0.0777  &  &0.0739   &0.0714    \\
    &    & MSE  &  &0.0036  &0.0027  &  &0.0065  &0.0061  &  &0.0055   &0.0051    \\ \hline 
\multicolumn{3}{c}{beta} &  & \multicolumn{2}{c}{$(\alpha,\beta)=(1,4)$} &  & \multicolumn{2}{c}{$(\alpha,\beta)=(5,3)$}&  & \multicolumn{2}{c}{$(\alpha,\beta)=(6,6)$}   \\ \cline{1-3} \cline{5-6} \cline{8-9} \cline{11-12} %\hline
 $t$  &  $n$ &  &  & $J^{w}_{n}$ & $J^{w}_{h}$ &  & $J^{w}_{n}$ & $J^{w}_{h}$ &  & $J^{w}_{n}$ & $J^{w}_{h}$    \\ \hline
0.01&  30& bias &  &0.0342  &0.0170  &  &0.1028  &0.0510  &  &0.0824   &0.0410    \\
    &    & MSE  &  &0.0012  &0.0004  &  &0.0111  &0.0035  &  &0.0824   &0.0023    \\
    &  50& bias &  &0.0326  &0.0170  &  &0.0963  &0.0491  &  &0.0773   &0.0397    \\ 
    &    & MSE  &  &0.0011  &0.0004  &  &0.0097  &0.0031  &  &0.0063   &0.0020    \\ 
    &    &  &  &  &  &  &  &  &  &   &    \\
0.10&  30& bias &  &0.0455  &0.0244  &  &0.1024  &0.0537  &  &0.0824   &0.0447    \\
    &    & MSE  &  &0.0022  &0.0008  &  &0.0112  &0.0040  &  &0.0073   &0.0027    \\
    &  50& bias &  &0.0426  &0.0230  &  &0.0946  &0.0512  &  &0.0771   &0.0426    \\ 
    &    & MSE  &  &0.0019  &0.0007  &  &0.0095  &0.0034  &  &0.0063   &0.0023    \\ 
    &    &  &  &  &  &  &  &  &  &   &    \\ 
0.30&  30& bias &  &0.0802  &0.0571  &  &0.1156  &0.0831  &  &0.0955   &0.0679    \\
    &    & MSE  &  &0.0072  &0.0039  &  &0.0149  &0.0082  &  &0.0101   &0.0055    \\
    &  50& bias &  &0.0726  &0.0517  &  &0.1054  &0.0756  &  &0.0880   &0.0633    \\
    &    & MSE  &  &0.0058  &0.0031  &  &0.0123  &0.0067  &  &0.0084   &0.0046    \\ \hline   
\end{tabular}
\end{table}
% % % % % % % % % % % % % % % % %
From Table \ref{tab-1}, it is seen that the MSE of $J^{w}_{n}(X,Y;t)$ and $J^{w}_{n}(X,Y;t)$ decreases as sample size $n$ increases. In both cases exponential and beta, the MSE of the estimator $J^{w}_{h}(X,Y;t)$ is less than that of $J^{w}_{n}(X,Y;t)$. The estimator based on the kernel method outperforms the one based on the empirical estimation of CDF. So, in practice, we recommend using approach based on the kernel estimations.
%######################################
{\color{red}
\section{Real data }
Here, we consider two real datasets to show the behavior of the estimator given in the previous part. \\
\textbf{First real dataset:}\\
The following dataset is given by Lee and Wang \cite{lee-wang-03} represented the remission times (in
months) for 128 patients with bladder cancer. This dataset is as\\
\begin{center}
2.09, 3.48, 6.94, 0.08, 4.87, 23.63, 8.66, 13.11, 3.52,
0.20, 2.23, 25.74, 4.98, 9.02, 13.29, 6.97, 2.26, 3.57, 0.40, 7.09, 5.06, 9.22, 13.80, 3.64, 0.50, 0.81, 2.46,
2.64, 5.09, 7.26, 9.47, 14.24, 25.82, 0.51, 2.54, 3.70, 5.17, 7.28, 9.74, 14.76, 26.31, 5.32, 2.62, 3.82, 12.07,
7.32, 14.77, 32.15, 10.06, 3.88, 5.32, 7.39, 10.34, 14.83, 34.26, 0.90, 2.69, 4.18, 5.34, 7.59, 10.66, 17.14,
36.66, 4.26, 15.96, 4.23, 1.05, 2.69, 8.65, 5.41, 10.75, 16.62, 7.62, 1.19, 2.75, 43.01, 11.25, 7.63, 5.41,
17.12, 1.26, 46.12, 2.83, 5.49, 4.33, 7.66, 3.36, 21.73, 22.69, 6.93, 4.50, 12.63, 2.07, 8.37, 79.05, 2.87, 5.62,
1.35, 11.64, 17.36, 7.87, 3.02, 4.34, 1.40, 7.93, 6.25, 5.71, 6.76, 12.02, 11.79, 18.10, 1.46, 2.02, 3.31, 4.51,
4.40, 5.85, 8.26, 6.54, 8.53, 12.03, 11.98, 19.13, 1.76, 20.28, 2.02, 3.36, 3.25.
\end{center}
For this dataset, we consider three candidate distributions. One of them is the log-logistic distribution with parameters $\alpha$ and $\lambda$ which has the following PDF and CDF in Equations \eqref{eq-ll-pdf} and \eqref{eq-ll-cdf} respectively, as
\begin{equation}\label{eq-ll-pdf}
g_{LL}(x;\alpha,\lambda)=\alpha \lambda^{-\alpha}x^{\alpha-1} \left( 1+(\frac{x}{\lambda})^\alpha \right)^{-2},
\end{equation}
and
\begin{equation}\label{eq-ll-cdf}
G_{LL}(x;\alpha,\lambda)=1-\left( 1+(\frac{x}{\lambda})^\alpha \right)^{-1}.
\end{equation}
We denote the log-logistic distribution by $\rm{LL}(\alpha,\lambda)$.
Recently, two generalization of the LL distribution were introduced. The alpha power transformed log-logistic (APLL$(\alpha,\lambda,a)$) was introduced by Aldahlan \cite{aldahlan-20} which has the following CDF in Equation \eqref{eq-apll-pdf} as
\begin{equation}\label{eq-apll-pdf}
F_{APLL}(x;\alpha,\lambda,a)=\frac{a^{G_{LL}(x;\alpha,\lambda)}-1}{a-1}.
\end{equation} 
Also, Alfaer et al. \cite{alfaer-21} introduced an extended version of log-logistic distribution (ExLL$(\alpha,\lambda,a)$) which has the following CDF in Equation \eqref{eq-exll-pdf} as
\begin{equation}\label{eq-exll-pdf}
F_{ExLL}(x;\alpha,\lambda,a)=1-
\left(\frac{1-G_{LL}(x;\alpha,\lambda)}{1-(1-a)G_{LL}(x;\alpha,\lambda)}\right)^a.
\end{equation}  
% % % % % % % % % % % % % % % % % % % % %
We fit the three above distributions to these data. 
The MLE of the parameters of the above distributions (LL, APLL and ExLL) are given in Table \ref{tab-mle}. 
\begin{table}[h]
\centering
\caption{MLE of parameters of proposed distributions.}\label{tab-mle}
\begin{tabular}{c|ccc}
\hline
parameter & LL & APLL & ExLL \\ \hline
$\hat{\alpha}$   &  $1.7251$    &   $1.7118$   &  $1.4276$   \\
$\hat{\lambda}$  &  $6.0898$    &   $4.9174$   &  $20.0321$  \\
$\hat{a}$        &   --         &   $2.0976$   &  $2.0701$   \\\hline

\end{tabular}
\end{table}
Also the Kolmogorov-Smirnov (K-S) statistics as well as its p-value of the above distributions are given in Table \ref{tab-k-s}.
\begin{table}[h]
\centering
\caption{K-S as well as p-value of proposed distributions.}\label{tab-k-s}
\begin{tabular}{c|ccc}
\hline
statistic & LL & APLL & ExLL \\ \hline
K-S    & 0.0399    &  0.0400   &  0.0351   \\
p-value&  0.9870   &   0.9866  &   0.9975  \\ \hline

\end{tabular}
\end{table} 
From Table \ref{tab-k-s}, all of the distributions can be fitted to this dataset at the type I error rate 0.05.
In the following, we examine two situations. In the first situation, we consider the LL as the actual distribution of the data and APLL as a distribution assigned by the experimenter. In the second situation, we consider the LL as the actual distribution of the data and ExLL as a distribution assigned by the experimenter.
For two situations, we compute $J^{w}_{n}(LL,\hat{F};t)$, $J^{w}_{h}(LL,\hat{F};t)$ and $J^{w}(LL,F_{0};t)$, where $F_{0}(.)$ can be either APLL or ExLL distribution.
The values of two estimators $J^{w}_{n}$ and $J^{w}_{h}$ and true value $J^{w}(LL,F_{0};t)$ are depicted in Figure \ref{fig-jwn-jw-2} as a function of $t$.

From Figure \ref{fig-jwn-jw-2}, it is seen that both $J^{w}_{n}$ and $J^{w}_{h}$ well estimate the value of $J^{w}$ in both plots. Indeed, in both cases, $J^{w}_{h}$ fits true value $J^{w}$ much better than $J^{w}_{n}$. So, we can use the estimator $J^{w}_{h}$ for our practical situations.
Also, as expected, the values of $J^{w}_{n}$ and $J^{w}_{h}$ as well as $J^{w}$ are decreasing function of $t$ in our two situations. 

\begin{figure}[h]
\centering
\includegraphics[width=6cm]{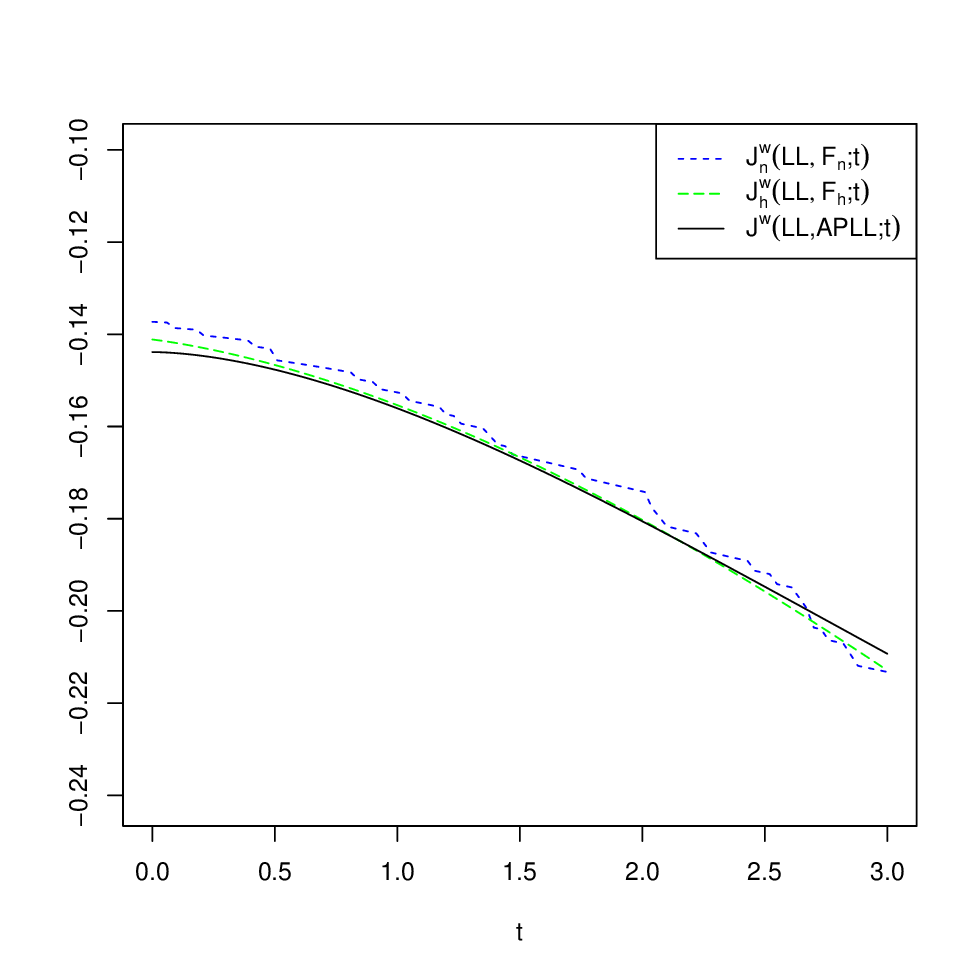}
\includegraphics[width=6cm]{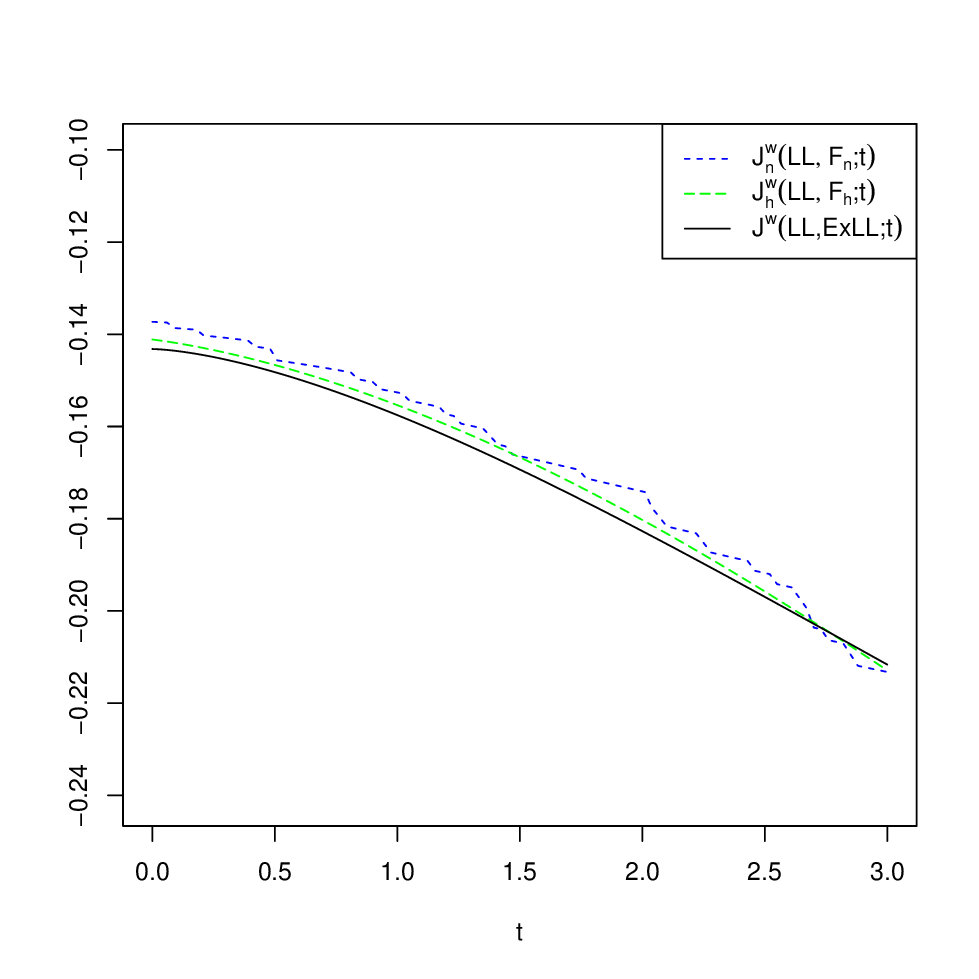}
\vspace{-0.5 cm}
\caption{The plot of $J^{w}_{n}$, $J^{w}_{h}$ and $J^{w}$: left plot is LL as an actual and APLL as a distribution assigned by the experimenter and right plot is LL as an actual and ExLL as a distribution assigned by the experimenter.}\label{fig-jwn-jw-2}
\end{figure}
Also, in the following Figure \ref{fig-apll-exll-comp}, we plot the values of $J^{w}(LL,F_{0};t)$ for both model APLL and ExLL as well as $J^{w}_{h}({LL,F_h};t)$ when LL is considered as the actual distribution of the data. Figure \ref{fig-apll-exll-comp} shows that when APLL distribution is assigned by the experimenter the obtained inaccuracy measure 
is lower than the case when the experimenter uses the ExLL model for these data. From Figure \ref{fig-apll-exll-comp}, we see that the estimated $J^{w}_{h}$ is closer to the $J^{w}{(LL,APLL;t)}$ than the $J^{w}{(LL,ExLL;t)}$ when we consider LL as the actual distribution for these data.
Therefore, APLL model provides a better approximation to these data when LL is the actual distribution in the sense that the SF $\bar{F}_{APLL}(x)$ is closer to the actual model $\bar{G}_{LL}(x)$ than SF $\bar{F}_{ExLL}(x)$.

\begin{figure}[h]
\centering
\includegraphics[width=10cm]{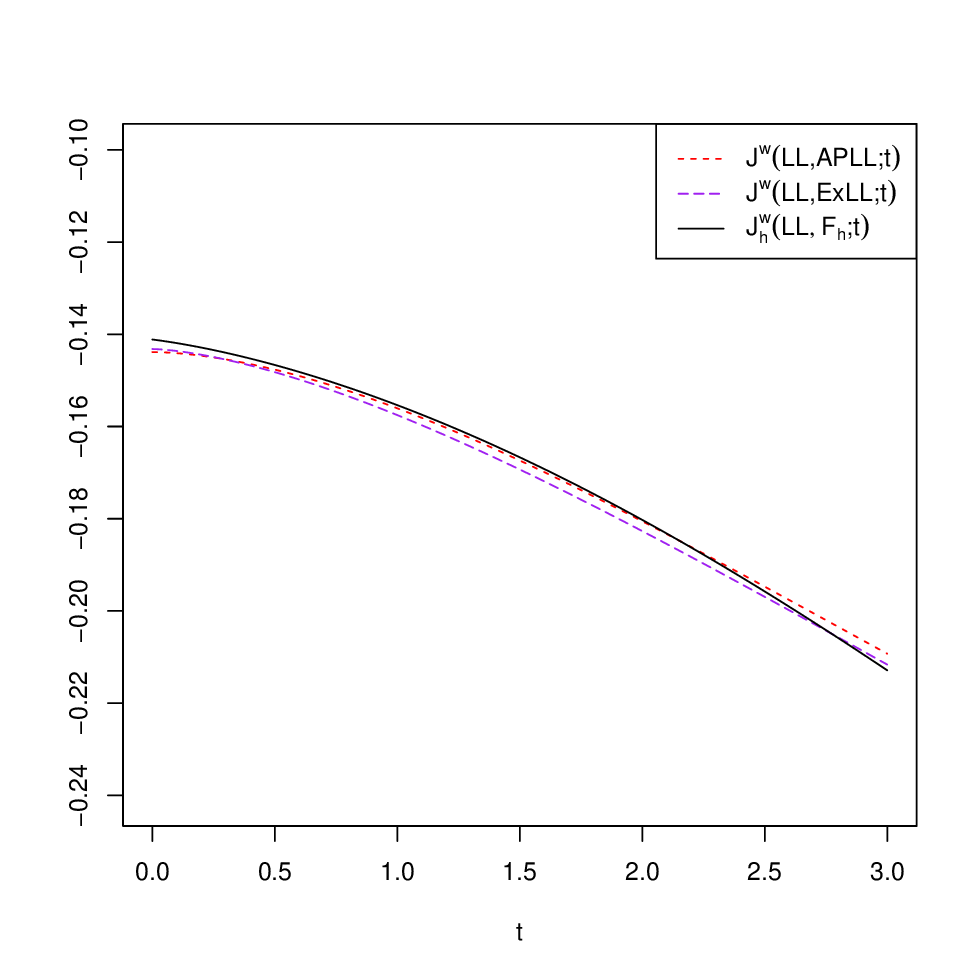}
\vspace{-0.6 cm}
\caption{The plot of $J^{w}_{h}$, $J^{w}{(LL,APLL;t)}$ and $J^{w}(LL,ExLL;t)$: LL model is used as an actual and APLL and ExLL models are assigned by the experimenter.}\label{fig-apll-exll-comp}
\end{figure}
% % % % % % % % % % % % % % % % % % %
% % % % % % % % % % % % % % % % % % % % % %
% % % % % % % % % % % % % % % % % % % % %
\textbf{Second real dataset:}\\
The following dataset is given by Bjerkedal \cite{bjerkedal-60} represented the the survival times (in days) of 72 guinea pigs
infected with virulent tubercle bacilli. This dataset is as
\begin{center}
0.1, 0.33, 0.44, 0.56, 0.59, 0.59, 0.72, 0.74, 0.92, 0.93, 0.96, 1, 1, 1.02, 1.05, 1.07, 1.07, 1.08,
1.08, 1.08, 1.09, 1.12, 1.13, 1.15, 1.16, 1.2, 1.21, 1.22, 1.22, 1.24, 1.3, 1.34, 1.36, 1.39, 1.44,
1.46, 1.53, 1.59, 1.6, 1.63, 1.68, 1.71, 1.72, 1.76, 1.83, 1.95, 1.96, 1.97, 2.02, 2.13, 2.15, 2.16,
2.22, 2.3, 2.31, 2.4, 2.45, 2.51, 2.53, 2.54, 2.78, 2.93, 3.27, 3.42, 3.47, 3.61, 4.02, 4.32, 4.58,
5.55, 2.54, 0.77.
\end{center}
For this dataset, three candidate distributions are considered. One of them is the Weibull distribution (WEI) with parameters $\lambda$ and $\gamma$ which has the following CDF in Equation \eqref{eq-ww-cdf} as
\begin{equation}\label{eq-ww-cdf}
G_{WEI}(x;\lambda,\gamma)=1-\exp^{-\lambda x^{\gamma}}.
\end{equation}
We denote this distribution by $\rm{WEI}(\lambda,\gamma)$.
The other two candidates are the gamma exponentiated-exponential (GEE)
(Ristic and Balakrishnan \cite{ris-bal-0}) and exponential-exponential geometric (EEG) (Rezaei et al. \cite{rez-et-al-0}) models. The GEE$(\lambda,\alpha,\theta)$ has the following PDF in Equation \eqref{eq-GEE-pdf} as
\begin{equation}\label{eq-GEE-pdf}
f_{GEE}(x;\lambda,\alpha,\theta)=\frac{\alpha \theta}{\Gamma(\lambda)}\exp^{-\theta x}(1-\exp^{-\theta x})^{\alpha-1}(-\alpha log(1-\exp^{-\theta x}))^{\lambda-1}.
\end{equation} 
Also, the EEG$(\alpha,\theta,p)$ has the following PDF in Equation \eqref{eq-EEG-pdf} as
\begin{equation}\label{eq-EEG-pdf}
f_{EEG}(x;\alpha,\theta,p)=\frac{\alpha \theta (1-p)\exp^{-\theta x}(1-\exp^{-\theta x})^{\alpha-1}}{(1-p+p(1-\exp^{-\theta x})^{\alpha})^2}.
\end{equation}  
% % % % % % % % % % % % % % % % % % % % %
We fit the three above distributions to these data. 
The MLE of the parameters of the above distributions (WEI, GEE and EEG) are given in Table \ref{tab-mle-ex2}. 
\begin{table}[h]
\centering
\caption{MLE of the parameters of the proposed distributions.}\label{tab-mle-ex2}
\begin{tabular}{c|ccc}
\hline
parameter       &   WEI        &      GEE     & EEG         \\ \hline
$\hat{\gamma}$  &  $1.7962$    &   --         &  --         \\
$\hat{\lambda}$ &  $0.2934$    &   $1.2899$   &  --         \\
$\hat{\alpha}$  &  --          &   $3.4676$   &  $3.5144$   \\ 
$\hat{\theta}$  &  --          &   $0.9118$   &  $1.1081$   \\ 
$\hat{p}$       &  --          &   --         &  $0.0343$   \\ \hline

\end{tabular}
\end{table}
Also the Kolmogorov-Smirnov (K-S) statistics as well as its p-value of the above distributions are given in Table \ref{tab-k-s-ex2}.
\begin{table}[h]
\centering
\caption{K-S as well as p-value of the proposed distributions.}\label{tab-k-s-ex2}
\begin{tabular}{c|ccc}
\hline
statistic &    WEI    &    GEE    &    EEG     \\ \hline
K-S       &   0.0982  & 0.0870    & 0.0883     \\
p-value   &   0.4902  & 0.6458    & 0.6284     \\ \hline

\end{tabular}
\end{table} 
From Table \ref{tab-k-s-ex2}, all of the distributions can be fitted to this dataset at the type I error rate 0.05.
As in the first real dataset, we examine two situations. In the first situation, we consider the GEE as the actual distribution of the data and WEI as a distribution assigned by the experimenter. In the second situation, we consider the GEE as the actual distribution of the data and EEG as a distribution assigned by the experimenter.
For two situations, we compute $J^{w}_{n}(GEE,\hat{F};t)$, $J^{w}_{h}(GEE,\hat{F};t)$ and $J^{w}(GEE,F_{0};t)$, where $F_{0}(.)$ can be either WEI or EEG distribution.
The values of the two estimators $J^{w}_{n}$ and $J^{w}_{h}$ and true value $J^{w}(GEE,F_{0};t)$ are depicted in Figure \ref{fig-jwn-jw-2-ex2} as a function of $t$.

From Figure \ref{fig-jwn-jw-2-ex2}, it is seen that $J^{w}_{h}$ fits true value $J^{w}$ much better than $J^{w}_{n}$. So, we can use the estimator $J^{w}_{h}$ for our practical situations.
%Also, as expected, the values of $J^{w}_{n}$ and $J^{w}_{h}$ as well as $J^{w}$ are decreasing function of $t$ in our two situations. 

\begin{figure}[h]
\centering
\includegraphics[width=6cm]{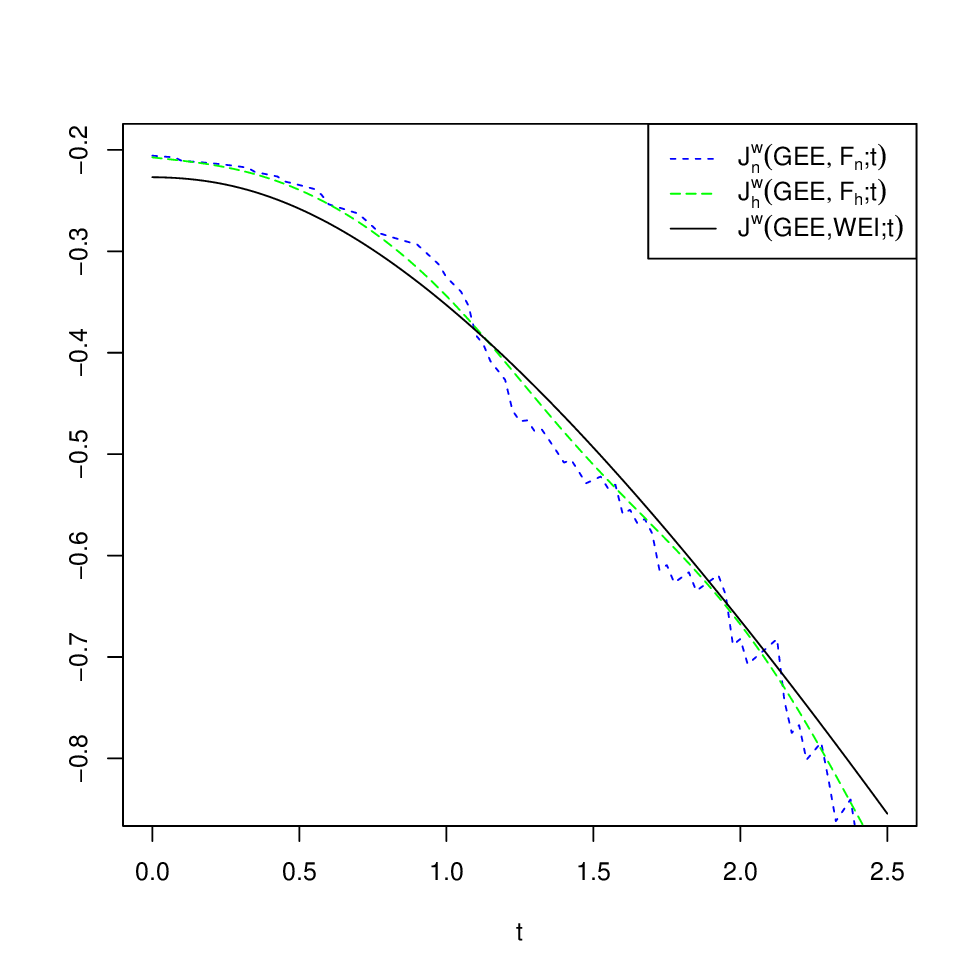}
\includegraphics[width=6cm]{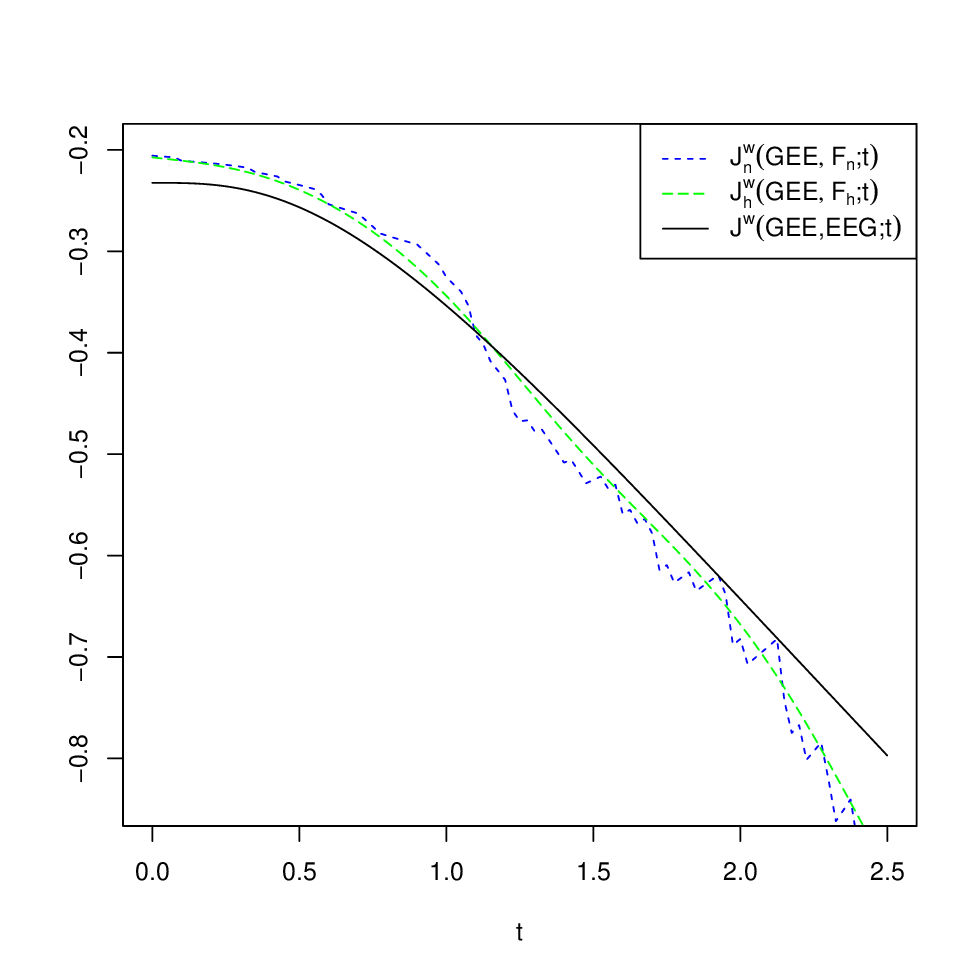}
\vspace{-0.5 cm}
\caption{The plot of $J^{w}_{n}$, $J^{w}_{h}$ and $J^{w}$: left plot is GEE as an actual and WEI as a distribution assigned by the experimenter and right plot is GEE as an actual and EEG as a distribution assigned by the experimenter.}\label{fig-jwn-jw-2-ex2}
\end{figure}
Similar to the first real dataset, in the following Figure \ref{fig-ww-eeg-comp}, we plot the values of $J^{w}(GEE,F_{0};t)$ for both model WEI and EEG as well as $J^{w}_{h}({GEE,F_h};t)$ when GEE is considered as the actual distribution of the data. Figure \ref{fig-ww-eeg-comp} shows that when WEI model is assigned by the experimenter the obtained inaccuracy measure 
is lower than the case when the experimenter uses EEG model for these data. From Figure \ref{fig-ww-eeg-comp}, we see that the estimated $J^{w}_{h}$ is closer to the $J^{w}{(GEE,WEI;t)}$ than the $J^{w}{(GEE,EGG;t)}$ when we consider GEE as the actual distribution for these data.

As we mentioned previously, the potential impact of WRJI extends to aid in model selection. So, organizations can make better decisions to select the optimized model in dynamic situations.

\begin{figure}[h]
\centering
\includegraphics[width=10cm]{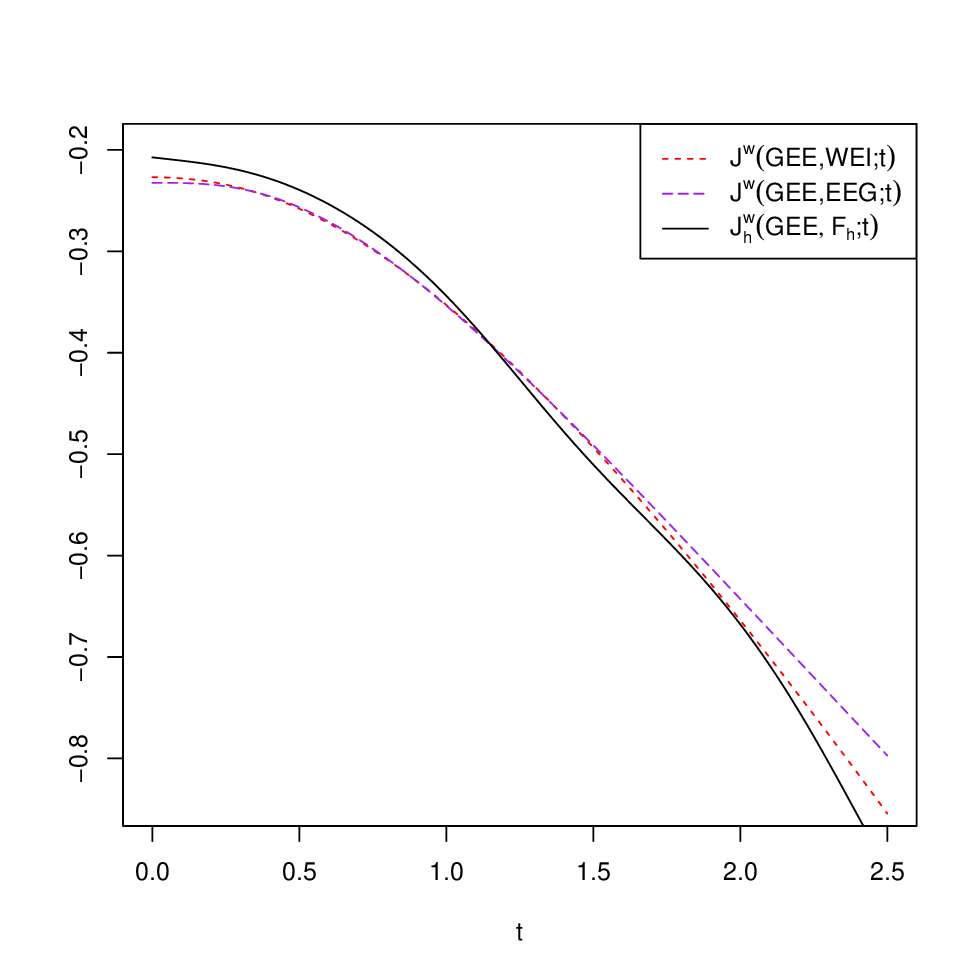}
\vspace{-0.6 cm}
\caption{The plot of $J^{w}_{h}$, $J^{w}{(GEE,WW;t)}$ and $J^{w}(GEE,EEG;t)$: GEE model is used as an actual and WW and EEG models are assigned by the experimenter.}\label{fig-ww-eeg-comp}
\end{figure}
}

\section{Discussion and Conclusion}\label{sec-conclusion}
The WRJI measure is a broadened concept of extropy that serves as a tool for measuring errors in experimental results. It combines an uncertainty measure and a discrimination measure between two distributions to quantify inaccuracies in statements about probabilities of events in an experiment. This measure is useful in statistical inference, estimation, and reliability studies for modeling lifetime data. In lifetime studies, data is often truncated, allowing for the extension of information-theoretic concepts to ordered situations and record values. By defining inaccuracy in terms of the WRJI measure, we can better characterize probability distributions and identify the most appropriate model for lifetime data. Traditional methods for finding the best model, such as goodness of fit procedures and probability plots, may fall short, making the WRJI measure a valuable tool in this context.
The introduction of WRJI in reliability modeling and decision-making is of practical importance in a variety of real-world situations. These measures offer valuable information on the accuracy and reliability of systems, enabling better decision-making. In the realm of reliability modeling, WRJI can be used to evaluate the performance of complex systems over time, pinpointing areas for improvement and enhancing system reliability. This is especially beneficial in industries like manufacturing, transportation, and energy, where system failures can have serious consequences. Additionally, it can support decision-making by providing a quantitative assessment of the reliability of different options or strategies. For instance, in finance, these measures can help assess investment portfolio performance and guide investment decisions. In health care, they can aid in evaluating the reliability of medical devices or treatment plans, empowering health care professionals to make more informed choices. Moreover, WRJI can be utilized for predictive maintenance, enabling organizations to proactively identify potential equipment failures and schedule maintenance activities efficiently. This proactive approach can reduce downtime, optimize resource allocation, and enhance operational efficiency overall. In essence, WRJI plays a crucial role in offering quantitative insights into reliability modeling and decision-making, enabling organizations to improve system performance, allocate resources effectively, and make well-informed decisions across various real-world scenarios. 
In this paper, by considering the concept of  residual extropy inaccuracy measure, its weighted  version was proposed. Under the assumption that the reference distribution $G$ and  true distribution $F$  satisfy the PHR model, it has been shown that the proposed measure determines the lifetime distribution uniquely. Moreover, upper and lower bounds and some inequalities concerning WRJI are determined. 
Two non-parametric estimators based on the kernel density estimation method for the proposed measures were also obtained. The performance of the estimators were also discussed using some simulation studies. A real data set was used for illustrating our estimators.
% % % % % % % % % % % % % % % % % % % % % % 
% % % % % % % % % % % % % % % % % % % % % % % %
%\section*{References}
\bibliographystyle{}
{}
% % % % % % % % % % % % % % % % % % % % % % % % % % %
% % % % % % % % % % % % % % % % % % % % % % % % % % %
% % % % % % % % % % % % % % % % % % % % % % % %
\end{document}